\documentclass[utf8]{FrontiersinHarvard} 

\usepackage{url,hyperref,lineno,microtype,subcaption}
\usepackage[onehalfspacing]{setspace}
\usepackage{xcolor}
\usepackage{placeins}
\usepackage{bm}
\usepackage[figuresright]{rotating}
\usepackage{booktabs}

\usepackage{xurl}

\def\keyFont{\fontsize{8}{11}\helveticabold }
\def\firstAuthorLast{Boyle {et~al.}} 
\def\Authors{G. J. Boyle\,$^{1,*}$, N. A. Garland\,$^{2,3,*}$, D. L. Muccignat\,$^{1}$, I. Simonovi\'{c}\,$^{4}$, \\ D. Bo{\v{s}}njakovi{\'c}\,$^{4}$, S. Dujko\,$^{4}$ and R. D. White\,$^{1}$}
\def\Address{$^{1}$College of Science and Engineering, James Cook University, Townsville, Australia \\
$^{2}$Queensland Quantum and Advanced Technologies Research Institute, Griffith University, Nathan, Australia \\
$^{3}$School of Environment and Science, Griffith University, Nathan, Australia \\
$^{4}$Institute of Physics, University of Belgrade, Belgrade, Serbia}
\def\corrAuthor{G. J. Boyle, N. A. Garland}

\def\corrEmail{gregory.boyle@jcu.edu.au, n.garland@griffith.edu.au}

\begin{document}
\twocolumn
\firstpage{1}

\title[Review of electron transport in noble liquids]{Review of the experimental and theoretical landscape of electron transport in noble liquids} 

\author[\firstAuthorLast ]{\Authors} 
\address{} 
\correspondance{} 

\extraAuth{}

\maketitle

\begin{abstract} 

We present a review of the current experimental and theoretical understanding of electron transport in noble liquids. Special attention is given to recent measurements that coincide with the development of time projection chambers (TPCs) using liquid xenon and argon as detector media. To enable transparent benchmarking of simulations and to facilitate the comparison between early studies and modern TPC data, we introduce a new open-access database of electron mobility and diffusion measurements. In particular, we emphasize the transition to large-scale detector designs which incorporate extended drift distances alongside improved purity control and field uniformity. On the theoretical side, we contrast empirical transport models with \textit{ab initio} approaches, highlighting our recent efforts to incorporate low-energy, liquid-specific scattering phenomena, including coherent scattering, polarization screening, and bulk potential modifications. While elastic transport has seen substantial theoretical progress, inelastic processes in liquids, including ionization, exciton formation and interband transitions, remain poorly understood due to the lack of experimental cross sections and validated models. We also discuss the applications and challenges of modeling scintillation, doped and mixture-liquid targets, and gas–liquid interface behavior, all of which are critical for the design and optimization of next-generation detectors.
 
\tiny
 \keyFont{ \section{Keywords:} electron transport, noble liquids, swarm experiments, time projection chambers, liquid xenon, liquid argon} 
\end{abstract}

\section{Introduction}\label{sec:Intro}

The transport of free electrons in dense fluids, particularly in liquefied noble gases such as argon and xenon, plays a central role in both fundamental physics and detector technology. These systems lie at the intersection of condensed matter, atomic physics, and high-energy physics, and underpin some of the most sensitive experimental searches for rare processes, including neutrino interactions and dark matter detection.

Liquid-phase noble gas detectors, especially those employing time projection chamber (TPC) designs, have become indispensable in astroparticle and neutrino physics. Simultaneous measurements of scintillation light and ionization charge signals allows precise event reconstruction. Recent large-scale experiments, including XENONnT~\citep{Aprile2019}, LZ~\citep{LUX-ZEPLINCollaboration2023}, PandaX~\citep{Liu2022}, ProtoDUNE~\citep{Abi2020}, MicroBooNE~\citep{microboone2021}, and ICARUS~\citep{TheICARUSCollaboration2004}, demonstrate this approach’s maturity, with tonne- to hundred-tonne-scale active masses now routine~\citep{EFCA2021}. For example, TPCs that use a LXe target have set world-leading limits on WIMP-nucleus cross-sections in the GeV/c$^2$ to TeV/c$^2$ mass range~\citep{Schumann_2019}.

In such detectors, particle interactions within the liquid target medium produce excited and ionized atoms. Scintillation photons offer a prompt time reference, while free electrons liberated by ionization are drifted under electric fields across the liquid-gas interface and are subsequently extracted. The fidelity of the signal reconstruction and hence the detector’s sensitivity depends critically on our understanding of how electrons move through the liquid. Properties such as drift velocity, diffusion, and recombination all influence the shape and timing of the charge signal.

To probe these transport properties, dedicated electron swarm experiments have been employed since the mid-20th century, many using time-of-flight (TOF) techniques. In these studies, a short electron pulse is produced (typically from a photocathode) and tracked as it drifts and diffuses under the influence of an electric field.  Some setups incorporate transverse magnetic fields to probe combined electric–magnetic field effects~\citep{Lamp1994}. Careful control of the thermodynamic state of the liquid, typically achieved by regulating temperature, pressure, and density, allows measurements over a broad range of physical conditions. Transport properties, such as mobility, characteristic energy, diffusion coefficients and the ionization rate, are then extracted from the evolution of the electron cloud. These properties feed directly into models for detector simulation and plasma behavior~\citep{MakaPetr06,Petrovic2009,RobsWhitHild17}, or can be used to extract underlying electron-atom scattering cross-sections via inverse swarm analysis~\citep{EngePhel63,EngePhelRisk64,Morg91b,StokEtal20,Muccignat_2024}.

Despite the widespread availability of gas-phase transport data, consolidated in databases like LXCat~\citep{PitcEtal17,Carbone_2021} and NIST~\citep{NIST}, corresponding liquid-phase data remain fragmented and incomplete.
In particular, few comprehensive datasets exist for liquids, especially in conditions relevant to modern TPCs. Furthermore, while simplified scaling laws often attempt to bridge gas and liquid regimes, this approach is generally insufficient due largely to the increased density effects that become important for electrons scattering in liquids compared to gases (see Section~\ref{sub:liquids}).

In 1965, Schnyders and collaborators questioned the applicability of gas-phase cross-sections in dense media~\citep{Schnyders1965}. Their subsequent work proposed that a more accurate description could be achieved by including interference effects via the liquid structure factor~\citep{Schnyders1966}. These insights, together with theoretical advancements by Cohen and Lekner~\citep{Lekner1967,CohenLekner1967}, laid the foundation for the modern understanding of electron transport in dense atomic fluids. The increased density introduces short-range order and coherent scattering effects, fundamentally altering the electron’s potential landscape. Early theoretical treatments addressed these issues by modifying the effective interaction potential to account for for the medium interparticle correlations~\citep{Lekner1967,CohenLekner1967,Schmidt1984,Sakai1982,SakiaNakamuraTagashira,Sakai2007,AtrazhevIakubov_1981,AtrazhevTimoshkin1996,AtrazhevTimoshkin1998}. Atrazhev and collaborators further demonstrated that, at low energies, cross-sections become nearly energy-independent and scale primarily with density~\citep{AtrazhevIakubov_1981,AtrazhevTimoshkin1996,AtrazhevTimoshkin1998}. In a later refinement, ~\cite{AtrazhevIakubovPogosov_1995} introduced a variable-phase-function method and muffin-tin potentials to incorporate liquid structure more realistically.

From an empirical perspective, Sakai and co-workers~\citep{SakiaNakamuraTagashira,Sakai2007} used a fitting procedure to adjust the momentum transfer cross-sections to match swarm data. Borghesani and co-workers~\citep{BorghesaniSantini1994,Borghesani2006,Borghesani2014} developed a heuristic for constructing an effective momentum transfer cross-section from modified gas phase cross-sections, thereby accurately predicting enhancements (or reductions) in the zero-field mobility. Sophisticated treatments that use a Green’s function method were employed by~\cite{BragliaDallacasa1982} to explore electron self-energy corrections, though such models are often limited by assumptions like linear response and near-equilibrium conditions. 

Our recent work aims to develop a unified \textit{ab initio} model for electron transport in noble liquids and dense gases, leveraging highly accurate electron-atom potentials benchmarked in the gas phase~\citep{BoyleEtal2015,BoyleEtal2016,WhiteEtal2018}. This framework avoids oversimplified potentials, e.g. the Buckingham potential, which neglect exchange interactions. The use of accurate potentials yields excellent agreement of transport coefficients calculated via a Boltzmann Equation (BE) solver with experimental data in the elastic scattering regime. However, extending this framework to incorporate inelastic processes, such as excitation and ionization, remains a frontier. Inelastic processes suppress coherence and introduce additional complexities, including threshold energy shifts and band structure formation~\citep{ReiningerEtal_1983,SimonovicEtal_2019}. Emerging tools such as machine learning may offer a valuable method to interpolate and validate cross-section models against liquid-phase data, but such efforts will require tight coupling between theory and precision measurement. 

Scintillation and luminescence underpin the operation of modern particle detectors, with signal formation driven by the creation of ion-electron pairs and the radiative decay of excimers formed in the liquid environment, and is thus intimately connected to ionization and excitation processes. Detector performance is highly sensitive to the microscopic details of these interactions, as well as to macroscopic effects such as impurity levels and the use of dopants, which can significantly alter light yield, spectral response, and charge transport characteristics. 

The macroscopic behavior of electron swarms in the detector medium is characterized by transport coefficients. A longstanding challenge in interpreting swarm data lies in the precise definition and measurement of these transport coefficients~\citep{Sakai_1977,TagashiraEtal77,Robson1984,Robson1986,KondoTagashira_1990,Robson1991}, and misunderstandings still persist. In the hydrodynamic regime, the space-time dependence of all macroscopic quantities can be projected onto the electron number density $n(r,t)$~\citep{KumaSkulRobs80}, and distinctions must be made between flux coefficients (those appearing in flux-gradient relations like Fick’s law) and bulk coefficients (those that describe the evolution of the entire cloud). These quantities can differ significantly in the presence of non-conservative processes such as ionization, attachment, or strong field gradients~\citep{Sakai_1977,TagashiraEtal77,Robson1984}. Although TPCs typically operate at low reduced electric field strengths in atomic liquids, there can be strong fields near the extraction region and the presence of impurities can selectively remove free electrons, and a formal kinetic or fluid model analysis is recommended.

In this work, we compile and critically examine the body of available measurements of electron transport in liquid xenon (LXe) and liquid argon (LAr). By making this dataset more accessible, we hope to support the broader community engaged in simulation, modeling, and detector design. Beyond the experimental survey, we also review the theoretical frameworks that have been developed to describe electron transport in atomic liquids. Special attention is given to recent advances, including \textit{ab initio} and semi-empirical models. New simulation tools, including emerging machine learning methods, offer new opportunities to inform detector development, particularly in regimes where direct measurements are difficult or incomplete.

In Section~\ref{sec:Swarms}, we review the current status of swarm experiments that measure electron mobility and diffusion in LAr and LXe, with a focus on results driven by TPC applications. Section~\ref{sec:Theory} examines the theoretical and simulation landscape, outlining how electron transport in noble liquids can be modeled from first principles and emphasizing key density effects, including elastic coherent scattering, bulk potential modifications, and the formation of excitation band-structures. In Section~\ref{sec:applications}, we focus on selected topical applications and experimental challenges in liquid detector development, covering topics such as scintillation modeling, the behavior of doped and mixed-liquid systems, and electron transport across gas-liquid interfaces. We conclude with a summary of current challenges and a forward-looking perspective on how modern modeling approaches and data-driven tools can shape the future of electron transport studies in noble liquid detectors.

\section{Noble Liquid Swarm Experiments}\label{sec:Swarms}

This section surveys key experimental efforts, as introduced in Section~\ref{sec:Intro}, to measure electron transport properties, specifically mobility and diffusion coefficients, in LAr and LXe. These measurements serve as essential benchmarks for validating the theoretical frameworks discussed in Section~\ref{sec:Theory}.

Early experiments generated electron swarms using radioactive sources placed near the cathode \citep{Williams1957}. Ionizing radiation from these sources produced electron–ion pairs \textit{in situ}, which were then accelerated through the liquid and collected at the anode~\citep{Davidson1950,Eibl1990}. While effective, these methods offered limited control over the initial distribution of electron energies and positions. Later techniques introduced more precise electron sources, such as photocathodes illuminated by pulsed lasers~\citep{Njoya2020} or discharge lamps~\citep{Eibl1990}, allowing for better-defined injection profiles. Additional methods include field emission from sharp metal electrodes immersed in the liquid~\citep{Halpern1967} and electron beams fired directly into the target medium~\citep{Miller1968}.

While the basic operating principles remain similar, modern large-scale detectors 
differ significantly from earlier laboratory-scale experiments. 
In particular, drift lengths have increased from millimeter–centimeter scales to distances of up to several meters \citep{EFCA2021}. This scaling introduces new challenges in maintaining field uniformity, ensuring liquid purity, and understanding long-range effects on charge transport.

Despite their utility, TOF experiments are subject to several sources of systematic uncertainty. One of the most important is the purity of the liquid. In the foundational experiments, impurities such as oxygen may have gone uncharacterized, potentially introducing electron attachment processes that distort transport measurements~\citep{Baudis2023}. In contrast, modern detectors benefit from advanced purification systems and use attachment rate measurements as a diagnostic for liquid purity.

Another challenge lies in accurately modeling the initial electron distribution. Techniques based on ionizing radiation tend to produce poorly defined spatial and energy profiles, which complicates direct comparison with simulations. Laser-based photoemission sources offer greater control but require precise alignment and timing. Similarly, improvements in cryogenic engineering now allow more accurate control of temperature, pressure, and density—properties that are essential for modeling the system’s phase and state variables. Earlier studies may lack this level of detail and, in such cases, approximations must be made to reconstruct the operating conditions and enable comparison with theory.

Additionally, assumptions of uniform electric fields may not hold near electrodes, where field distortions can affect drift and diffusion. This can introduce bias when interpreting TOF data, particularly for short drift lengths or near-surface effects. Since the TOF method averages transport properties over the full drift region, it is inherently limited in its ability to resolve spatial variations or localized inhomogeneities in the liquid.

Nonetheless, TOF remains one of the most powerful and widely used techniques for probing electron transport in atomic liquids. When conducted under well-characterized conditions, these experiments provide critical data for developing and validating theoretical models, as well as informing the design and operation of noble-liquid-based detectors.

\subsection{Select measurements of electrons in LXe and LAr}\label{sub:argon}

As noted in Section~\ref{sec:Intro}, tabulated transport coefficients and scattering cross-sections play a vital role in plasma modeling and in enabling transparent, quantitative comparisons across different experimental studies. Over the past decade, the expansion of experimental datasets has also enabled data-driven approaches, most notably deep learning techniques, to infer electron scattering cross-sections from measured transport properties~\citep{StokEtal20,Jetly2021,Muccignat_2024}. To date, most databases (including LXCat and NIST) have focused primarily on electron transport in the gas phase. However, recent developments have begun to broaden this scope. For example, LXCat has announced an updated version of its database that will incorporate new states of matter and broader classes of incident particles~\citep{Boer2023}.

In this section, we present representative data on the electron mobility and diffusion coefficients in LXe and LAr over a range of experimental conditions. The full dataset is available on GitHub\footnote{Liquid Transport Data - \url{www.github.com/jcu-transport-physics/liquid-transport-data}} and is structured in a format similar to LXCat. While CSV formats were initially used, we are exploring alternative text-based formats to better support transparency and long-term maintainability. 
A complete list of the data in the repository is available in Appendix~\ref{appendix:data}. 

These representative samples highlight both the range of experimental behavior observed and the degree of consistency between early pre-2000 measurements and more recent studies motivated by large-scale liquid TPC development. Because published transport data are typically reported as functions of the absolute electric field at fixed temperatures and pressures, we have converted the data into reduced quantities (e.g., $E/N$) using reported liquid densities where available. In cases where the density is not given, we estimate it along the saturated liquid line using interpolated data from the NIST Chemistry WebBook\footnote{NIST Standard Reference Database 69: NIST Chemistry WebBook - \url{https://doi.org/10.18434/T4D303}} database of Thermophysical Properties of Fluid Systems~\citep{NIST}.

Transport in these systems is often modeled using the drift-diffusion equation~\citep{HuxlCrom74}, which requires knowledge of drift velocity (mobility), swarm broadening (diffusion), and net particle production or loss (ionization or attachment). In noble liquids, at low fields, electron impact ionization does not occur and attachment is negligible in pure media. However, even trace impurities such as oxygen or water vapor can introduce attachment pathways, emphasizing the importance of high-purity conditions in experiments. For TPC applications, high mobility and low diffusion are especially desirable, motivating the use of liquid-phase detectors.

\begin{figure}[t]
    \centering
    \includegraphics[width=0.95\linewidth]{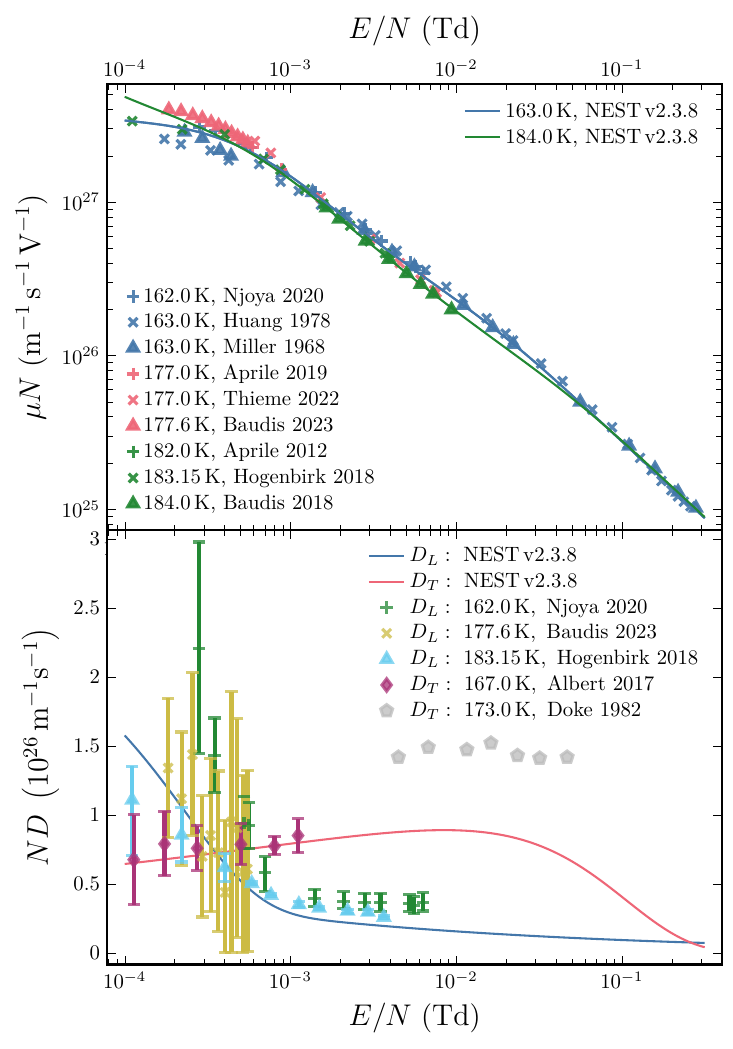}
    \caption{Top: Experimental reduced mobility ($\mu N$) as a function of reduced electric field ($E/N$) for LXe restricted to three groups of temperatures. Each color represents a different temperature grouping. Solid lines are the empirical curves from NEST~\citep{szydagis2022review}. Bottom: Experimental reduced diffusion coefficients ($ND_L$ and $ND_T$) as a function of reduced electric field from select authors for LXe. Solid lines are the empirical curves from NEST.} 
    \label{fig:Xe_reduced2}
\end{figure}

The top panel of Figure~\ref{fig:Xe_reduced2} shows the reduced electron mobility in LXe as a function of reduced electric field, spanning more than 50 years of measurements. The data are grouped by background temperature and compared to empirical fits from the NEST simulation framework~\citep{szydagis2022review,szydagis2025review}.

Recent measurements have largely been limited to low reduced fields ($E/N<0.01$\,Td), making high-field behavior less well characterized. The TPC data suggest higher mobilities at similar conditions, e.g., when comparing the data of \cite{Huang1978,Miller1968}, and that of \cite{Njoya2020}. This may reflect differences in experimental design, particularly liquid purity and field uniformity. The mobility curves at 177\,K and 183\,K exhibit slightly different slopes compared to those at 163\,K, potentially pointing to temperature-dependent structural changes or impurity effects. The NEST simulation results align well with all measurements, favoring the more recent measurements at low fields, although this agreement likely reflects the empirical basis of the model. As~\cite{Baudis2023} notes, impurities such as water vapor and hydrocarbons can serve as additional energy-loss channels, leading to higher mobility and suppressed diffusion. This may help explain the elevated drift velocities reported in early studies such as~\cite{Gushchin1982}, where purification methods were not documented.

Figure~\ref{fig:Xe_reduced2} also presents reduced longitudinal and transverse diffusion coefficients in LXe, in the bottom panel. Although uncertainties are significant, the longitudinal diffusion coefficients decrease with increasing field strength, while the transverse diffusion coefficients appear largely field-independent within each dataset. The trends in the longitudinal diffusion are broadly consistent with NEST simulations, though further high-precision measurements would help constrain the full field dependence and reduce uncertainty. 
The NEST curve for the transverse diffusion coefficient has been determined from the experimental measurements of \cite{Albert2017}, but underestimates the measurements at higher fields from \cite{Doke1982} (which has an estimated uncertainty of approximately $\pm 10$\% \citep{Doke1981}). Clearly, a single model consistent with the full range of LXe transport coefficients is still elusive. 

\begin{figure}[t]
    \includegraphics[width=0.95\linewidth]{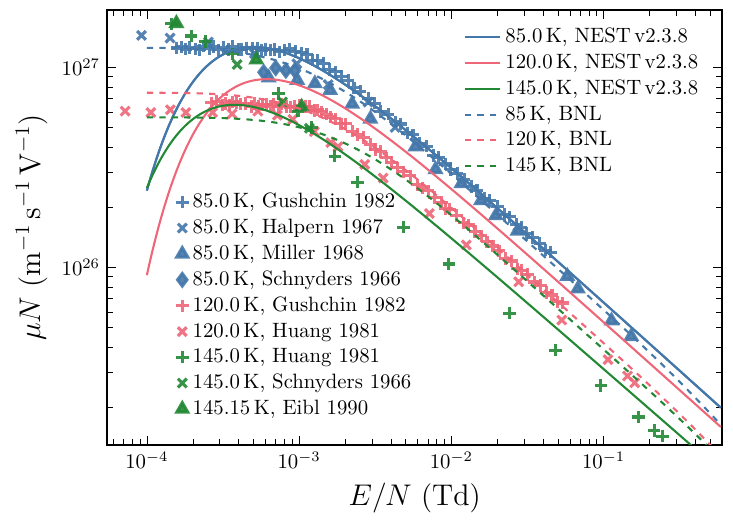}
    \caption{Experimental reduced mobility ($\mu N$) as a function of reduced electric field $E/N$ for LAr, restricted to three groups of temperatures. Each color represents a different temperature grouping. Solid lines are the empirical curves from NEST \citep{szydagis2022review}. Dashed lines are the empirical curves from BNL \citep{Li2016}}
    \label{fig:Ar_reduced2}
\end{figure}

Figure~\ref{fig:Ar_reduced2} shows the reduced mobility of electrons in LAr, with data grouped by temperature over the range of 85 K to 150 K. The results show considerable variation in both plateau values and field dependence, reflecting the underlying temperature sensitivity of the medium’s structure. In particular, the nonlinear temperature dependence of the pair correlation function and the static structure factor plays a key role at low fields.

Within each temperature group, measurements are generally consistent, with the exception of the early results reported by~\cite{Gushchin1982}, which again appear anomalously high when compared to recent experiments. These discrepancies are likely attributable to uncharacterized impurities.

Temperature clearly plays an important role in determining transport behavior, not just through thermal velocity distributions, but also via structural correlations in the liquid. This sensitivity presents challenges for empirical modeling. The NEST profiles do a sensible job of representing the mobility data at reduced fields above 0.001 Td, but break down at low fields. The Brookhaven National Laboratory (BNL) tool’s empirical fit \citep{Li2016} performs reasonably well at 85 K and 120 K but fails to reproduce data at 145 K. This highlights a broader problem: empirical models often lack predictive power outside the narrow conditions for which they were calibrated, reinforcing the need for \textit{ab initio} approaches that incorporate medium structure and density effects explicitly. 

Although LAr and LXe are the most studied noble liquids, other systems have also been investigated. Measurements in liquid krypton are broadly consistent with LAr and LXe results~\citep{Nishikawa2007,Schnyders1966}, showing intermediate behavior. In contrast, transport in neon and helium shows marked differences due to their unique electronic properties. Neon and helium have low polarizability and exhibit positive scattering lengths due to Pauli exclusion, resulting in fundamentally different scattering dynamics~\citep{Bruschi1972,Sakai1982,Borghesani1988,Hernandez1991,Schmidt2003,Borghesani2021-fk}. These liquids support the formation of electron bubbles, i.e., quasi-stable low-density cavities, due to strong short-range repulsive forces \citep{Hernandez1991}. As a result, mobility in neon and helium is often lower in the liquid than in the gas phase. By contrast, in argon and xenon, attractive interactions promote electron localization at atomic sites, preventing bubble formation and leading to higher liquid-phase mobility \citep{Borghesani1992a,Borghesani2006}. Electron localization and self-trapping is discussed further in Section~\ref{sub:trapping}. 

\FloatBarrier

\section{Theory and Simulation}\label{sec:Theory}

The transport of charged particles through gases and liquids is most rigorously described by Boltzmann's kinetic equation (BE). Since its introduction in the 19th century~\citep{Boltzmann1872}, the BE has remained the gold standard for understanding the statistical behavior of charged particles in a wide range of systems~\citep{CoheThir73,Boyle_etal23}.

However, the BE approach is most effective in idealized scenarios i.e., those with simplified geometries, uniform media, and well-behaved boundary conditions. Real-world systems, particularly those relevant to detector physics, often involve complexities such as sharp interfaces, spatially varying fields, and secondary particle production. These features can make solutions of the deterministic BE intractable.

To address these challenges, stochastic methods, most notably Monte Carlo (MC) simulations \citep{Skullerud1968,Park2023,Beever2024}, have been widely adopted. Modern MC techniques, along with particle-in-cell or hybrid schemes, provide a powerful and flexible alternative to deterministic modeling~\citep{Tskhakaya2007}. With the increasing availability of computational resources, these methods have become the preferred tool for exploring systems with complex spatial structures, non-linear effects, and multiple interacting particles. Previous
studies have demonstrated the utility of MC
simulations for modeling electron transport in both
gas and liquid media~\citep{Emfietzoglou2003,Champion2012,Petrovic2014,Tattersall2015,Emfietzoglou2017,Mehnaz2020,Beever2024}.

In the context of particle detectors, one of the most widely used MC-based tools is the Noble Element Simulation Technique (NEST) software~\citep{Szydagis2011,Szydagis2021,szydagis2025review}. NEST has been extensively validated against experimental data and is frequently employed to simulate detector responses across a wide range of energies and electric field strengths. Its phenomenological approach enables practical modeling of scintillation and ionization signals in both gaseous and liquid systems. However, as noted in~\cite{szydagis2025review}, NEST relies heavily on empirical parameterizations which may introduce significant uncertainties when extrapolating to new energy regimes or electric field conditions where data is sparse or unavailable. Integrating \textit{ab initio} scattering cross-sections and transport models into frameworks like NEST would represent an important step forward, reducing reliance on empirical tuning and enhancing predictive power.

While this section primarily focuses on the application of the BE to noble gas and liquid systems, many of the underlying scattering processes discussed, such as interference effects and potential modifications, are equally relevant to MC simulations. In practice, the most robust modeling strategies will combine the strengths of both deterministic and stochastic methods. Together, these approaches provide a complementary toolkit for the accurate simulation of electron transport in complex detector environments.

\subsection{Multi-term solution of the Boltzmann kinetic equation}\label{sub:Boltz}

The behavior of electrons in gases and liquids under the influence of an external electric field $\bm{E}$ is governed by the BE, which describes the evolution of the phase-space distribution function $f(\bm{r},\bm{v},t)$. This function represents the probability density for finding an electron with velocity $\bm{v}$ at position $\bm{r}$ and time $t$. The BE is~\citep{Boltzmann1872,RobsWhitHild17}: 
\begin{align}
    \left(\frac{\partial}{\partial t}+\bm{v}\!\cdot\!\nabla + \frac{e\bm{E}}{m_e}\!\cdot\!\frac{\partial}{\partial\bm{v}}\right)\!f(\bm{r},\bm{v},t)\! =\! -J(f), \label{eq:Boltz}
\end{align}
where $e$ and $m_e$ are the electron charge and mass, and $
J(f)$ is the collision operator describing interactions between electrons and the background medium.

In the dilute (or swarm) limit, where the electron population is sparse and does not perturb the background gas, the collision operator remains linear and can be decomposed into contributions from various interaction processes:
\begin{align}
    J(f) = J_{el}(f) + J_{ex}(f) + J_{at}(f) + J_{io}(f) + \cdots.
\end{align}
with $J_{el}(f)$, $J_{ex}(f)$, $J_{at}(f)$, and $J_{io}(f)$ representing elastic scattering, excitation, attachment, and ionization, respectively. Boltzmann’s original formulation only included elastic processes \citep{Boltzmann1872}, but subsequent extensions have incorporated inelastic collisions, particle creation and loss, quantum effects, and even relativistic or condensed-phase considerations (see \cite{RobsWhitHild17} for a review).

For systems involving electrons it is common to simplify the collision operator using a mass-ratio expansion in $m_e/(m_e+m_0)$, where $m_0$ is the mass of the neutral target~\citep{Kumar67,RobsWhitHild17}. For light projectiles, only the first-order term is typically required, which simplifies the analysis significantly~\citep{ChapCowl70,FrosPhel62,KumaSkulRobs80}.

Number-conserving processes, such as elastic collisions and electronic excitations, are described using the semi-classical Wang Chang–Uhlenbeck–de Boer operator~\citep{WangEtal64}. For excitation processes, this is expressed as:
\begin{align}
    &J_{ex}(f)\! = \sum_{j, k}\!\int\! d\hat{\bm{g}}\ d\hat{\bm{v}_0}\  g\sigma(j,k;g,\chi) \nonumber \\
    &\ \ \ \ \ \ \times\!\left[f(\bm{r},\!\bm{v},\!t)F_{0j}(\bm{v}_0) \!-\! f(\bm{r},\!\bm{v}',\!t)F_{0j}(\bm{v}_0')\right],
\end{align}
where $j$ and $k$ label the internal states of the background atoms, $\sigma(j,k;g,\chi)$ is the differential cross-section for the transition $(j,\bm{v},\bm{v}_0)\rightarrow (k,\bm{v}',\bm{v}_0')$, and $F_{0j}(\bm{v}_0)$ is the Maxwellian distribution for background atoms in state $j$. Energy conservation links the pre- and post-collision velocities via $\frac{1}{2}m_eg^2 + U_j = \frac{1}{2}m_eg'^2 + U_k$, where $U_j$ and $U_k$ are the internal energies of the respective states. When $j=k$, the collision represents an elastic scattering event. Modifications of the elastic scattering process in a dense gas or liquid medium, such as coherent scattering, are discussed in Section~\ref{sub:liquids}.

In contrast, non-conserving processes like electron attachment and ionization remove or create electrons, respectively. Attachment processes are described by:
\begin{align}
    J_{at}(f)\! =\! \sum_{j}\! f(\bm{r},\!\bm{v},\!t)\!\!\int\!\! d\bm{v}_0\  g\sigma_{at}(j;\!g) F_{0j}(\bm{v}_0), 
\end{align}
where $\sigma_{at}(j;g)$ is the attachment cross-section for the $j$th attachment process. Ionization is represented as:
\begin{align}
    &J_{io}(f) = \sum_{j}n_{0j}v\sigma_{io}(j;v)f(\bm{r},\!\bm{v},\!t) \nonumber \\
    &- \!2n_{0j}\!\int\! d\bm{v}'\  v'\sigma_{io}(j;v') B(\bm{v},\!\bm{v}';\!j')f(\bm{r},\!\bm{v}',\!t),
\end{align} 
where $\sigma_{io}(j;v)$  is the ionization cross-section for the $j$th process, and $B(\bm{v},\bm{v}';j')$ is the probability distribution for the velocities of the outgoing electrons. Although ionization is fundamentally a three-body process, it is typically assumed that the heavy neutral remains stationary (valid to zeroth order in the mass ratio) so that energy and momentum are shared only between the incident and ejected electrons.

To solve the BE, a common strategy is to expand the distribution function in spherical harmonics, $Y_m^{[l]}$~\citep{NessRobs86,RobsNess86,RobsWhitHild17}: 
\begin{align}
    f(\bm{r},\bm{v},t) = \sum_{l=0}^{l_{\textrm{max}}}\sum_{m=-l}^lf_m^{(l)}(\bm{r},v,t)Y_m^{[l]}(\bm{\hat{c}}).\label{eq:spherharm}
\end{align}
In systems with a preferred direction, such as those with a uniform electric field, this simplifies to a Legendre polynomial $P_l$ expansion:
\begin{align}
    f(\bm{r},\bm{v},t) = \sum_{l=0}^{l_{\textrm{max}}}f_l(\bm{r},v,t)P_l(\cos{\theta}),\label{eq:legendre}
\end{align}
with $\theta$ being the angle between the velocity vector and the electric field. Although the two-term approximation ($l_{max}=1$) has been widely used for electrons, its limitations are well established~\citep{RobsNess86,WhitEtal03,Petrovic2009}, and convergence must be tested by systematically increasing $l_{max}$.

Various representations of the speed-space dependence of the coefficients in Equations~\eqref{eq:spherharm}-\eqref{eq:legendre} have been previously employed, including Sonine polynomials~\citep{Burnett35,Burnett36}, cubic splines~\citep{PitcNeilRumb81,Penetrante1985}, finite volume~\citep{Boyl17}, and finite difference schemes~\cite{MaedMaka94,LoffWink96,SigeWink96,TrunBonaNeca06,Stephens2018b}. The combination of spherical harmonics and Sonine polynomials yields the well-known Burnett functions \citep{Kumar66,Burnett35,Burnett36}, which have played a critical role in the development of a unified solution to Boltzmann’s equation across all mass ratios.

Regardless of the exact speed-space representation, substituting the angular expansion into the BE yields a hierarchy of coupled differential equations for the expansion coefficients $f_m^{(l)}(\bm{r},v, t)$. In the spatially homogeneous steady-state case, these take the form:
\begin{align}
    J^l(f_m^{(l)}) - \frac{eE}{m_e}\Bigg(\frac{l+1}{2l+3}\left[\frac{d}{dv}+\frac{l+2}{v}\right]f_m^{(l+1)}& \nonumber \\
    -\frac{l}{2l-1}\left[\frac{d}{dv}-\frac{l-1}{v}\right]f_m^{(l-1)}\Bigg) = 0&,
\end{align}
where $J^l$ denotes the spherical harmonic component of the collision operator.

Once these coefficients are known, key physical observables can be calculated. The electron number density $n$ is given by:
\begin{align}
    n &= 4\pi \int_0^\infty dv\  v^2f_0^{(0)}(v),
\end{align}
and is typically normalized to unity. The flux drift velocity $W$ is obtained from:
\begin{align}
    W = \frac{4\pi}{3}\int_0^\infty dv\ v^3f_0^{(1)}(v),
\end{align} 
while the mean energy $\epsilon$ is:
\begin{align}
    \epsilon &=  4\pi \int_0^\infty dv  \frac{1}{2}m_ev^4f_0^{(0)}(v).
\end{align} 
Finally, the effective ionization coefficient (normalized to the neutral density $N$) is expressed as:
\begin{align}
    \frac{R}{N} &=  4\pi\! \int_0^\infty\!\!dv \  v^3\left[\sigma_{gain}(v)\!-\!\sigma_{loss}(v)\right]f_0^{(0)}(v),
\end{align}
where $\sigma_{gain}$ and $\sigma_{loss}$ include all relevant cross-sections for particle-creating and particle-depleting processes, respectively.

To extend these results beyond spatial homogeneity, the distribution functions $f_m^{(l)}(\bm{r},v,t)$ can be expanded in terms of spatial and temporal gradients. This formalism connects directly to hydrodynamic transport coefficients, including bulk and flux mobilities and diffusion tensors, thus providing a bridge between microscopic kinetics and macroscopic fluid descriptions~\citep{RobsWhitHild17,RobsNess86,NessRobs86,WhitEtal09}. 

There are often multiple common ways of presenting the same physical phenomena, which can depend on the exact type of measurement undertaken. For instance, the electron mobility $\mu = W/E$ (or reduced electron mobility $N\mu$) is often tabulated instead of the drift velocity, while the ionization coefficient $\alpha \approx R/W$ is measured in a steady-state Townsend experiment~\cite{HuxlCrom74,Dujko2008}. The transverse diffusion coefficient $D_T$ is routinely combined with the mobility $\mu$ and presented as the characteristic energy, $\tfrac{3}{2}D_T/\mu$.

\subsection{Electron swarms in atomic liquids}\label{sub:liquids}

\subsubsection{Interference effects}\label{sub:interference}

In dense media such as liquid noble gases, the transport of electrons differs significantly from that in dilute gases due to the emergence of quantum interference effects. These effects arise when the de Broglie wavelength of an electron becomes comparable to the average spacing between atoms in the medium. Under such conditions, scattering can no longer be treated as isolated binary collisions. Instead, electrons scatter coherently from multiple, spatially and temporally correlated atoms, leading to interference phenomena that are absent in low-density environments.

At intermediate densities, where the electron’s mean free path remains much larger than its de Broglie wavelength, coherent scattering can be modeled under the single-scatterer approximation. In this regime, the medium is treated as a single extended scattering center with spatial correlations encoded through the structure of the surrounding atoms~\citep{CohenLekner1967}. This coherent elastic scattering effectively modifies the angular dependence of the collision process but does not apply to inelastic interactions. Processes such as excitation and ionization involve localized energy transfer and thus disrupt the phase coherence required for interference. As a result, interference effects are considered only in elastic channels.

As density increases further, the electron’s mean free path can approach its thermal de Broglie wavelength. As described in the review by \cite{Borghesani2006}, in this regime the electron wavefunction may interfere with itself along time-reversed paths that scatter off neighboring atoms. This quantum self-interference leads to enhanced elastic backscattering and contributes to the suppression of electron mobility. At very high densities, these effects can become strong enough to induce a mobility edge~\citep{Polischuk1984}, signifying a transition to localized transport reminiscent of Anderson localization in disordered solids. Such behavior marks the onset of a qualitatively different transport regime, where interference effects dominate over collisional scattering. 

In typical noble-liquid detector environments, the atomic number density is $~1-2\times10^{28}$ m$^{-3}$, corresponding to Wigner–Seitz cell radii of approximately $5$a$_0$. The de Broglie wavelength of electrons with energies below 10 eV is greater than 7a$_0$, necessitating the inclusion of interference effects.

To incorporate these effects quantitatively, the elastic collision operator in the BE is modified through structure-dependent corrections. The Legendre components of the collision operator accounting for coherent scattering are given by:
\begin{align}
    J_0 (\Phi_0)\! &=\! \frac{m_e}{m_0v^2}\!\left[v\nu_1(v)\!\left(vf_l \!+\! \frac{k_bT}{m_e}\!\frac{d}{dv}f_0\right)\right]\!, \label{eq:J0} \\
    J_l (\Phi_l) &= \tilde{\nu}_l(v) f_l,\ \ \ \ \text{for } l \geq 1, \label{eq:Jl} 
\end{align} 
where $m_0$ and $T$ are the mass and temperature of the background neutrals. The binary collision frequency $\nu_l(v)$ is given by,
\begin{align}
    \frac{\nu_l(v)}{N} &= 2\pi v\!\!\int_0^{2\pi}\!\!\sigma(v,\theta)[1-P_l(\cos\theta)]\sin\theta d\theta,\label{eq:nul}
\end{align}
and the structure-modified collision frequency $\tilde{\nu}_l(v)$ is:
\begin{align}
    \frac{\tilde{\nu}_l(v)}{N} &= 2\pi v\!\!\int_0^{2\pi}\!\!\Sigma(v,\theta)[1-P_l(\cos\theta)]\sin\theta d\theta,\label{eq:nultilde}
\end{align}
where $\Sigma(v,\theta)$ represents the structure modified differential cross-section:
\begin{align}
    \Sigma(v,\theta)
    &=  \sigma(v,\theta)S\left(\frac{2m_ev}{\hbar}\sin{\frac{\theta}{2}}\right), \label{eq:DCS}
\end{align}
which incorporates interference effects via the static structure factor $S(q)$, with $q=\frac{2mv}{\hbar}\sin{\left(\frac{\theta}{2}\right)}$ representing the momentum-transfer. The structure factor $S(q)$ encodes spatial correlations among atoms and can be obtained from the Fourier transform of the pair correlation function~\citep{HansenMcDonald1973}. In the dilute limit, $S(q)\rightarrow 1$, and the expressions reduce to those for uncorrelated binary scattering. The extension to liquid mixtures is discussed further in Section~\ref{sub:mixtures}.

\subsubsection{Modifications to the scattering potential}\label{sub:potentials}

\begin{figure}[t]
\begin{center}
\includegraphics[width=6.0cm]{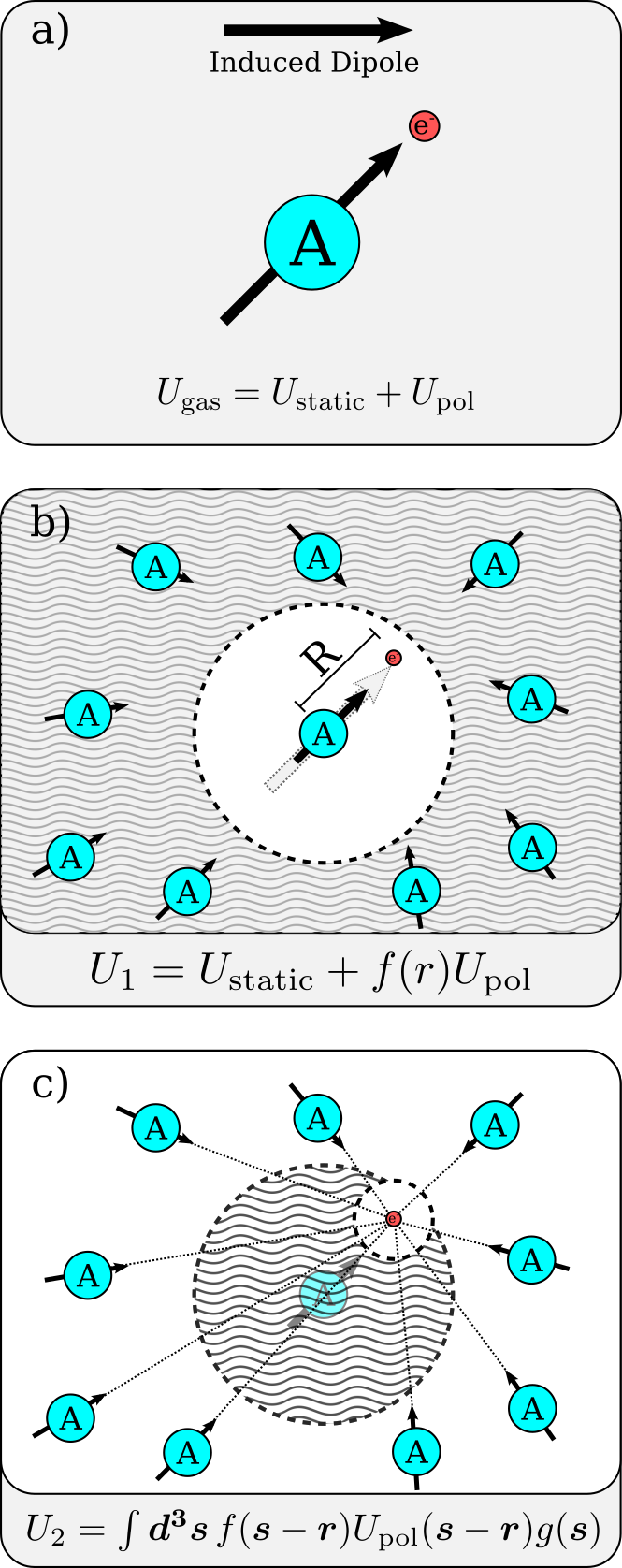}
\end{center}
\caption{Schematic representation of the various components of the
screening of the electron–atom potential in a liquid environment. (a) Gas phase potential is a combination of the static interaction potential $U_{static}$ and the polarisation component $U_{pol}$. (b) Interaction potential $U_1$ associated with the ‘focus atom’ is a combination of $U_{static}$ and the polarisation potential screened by the surrounding atoms. (c) Interaction potential $U_2$ associated with the surrounding atoms. Here $r$ is the position of the electron and $s$ the position of surrounding atoms.(Source: Reproduced from~\cite{WhiteEtal2018}, with the permission of IOP Publishing.)}\label{fig:liquideffects}
\end{figure}

In addition to interference effects, the electron scattering process in dense fluids is profoundly influenced by the collective response of the surrounding medium. Two primary contributions must be considered: first, the screening of the long-range polarization potential of the focus atom by neighboring atoms, and second, the direct contribution of those surrounding atoms to the total potential experienced by the electron. Figure~\ref{fig:liquideffects} depicts these contributions. These effects fundamentally alter the effective interaction potential and, consequently, the transport properties of electrons in liquids. 

In an isolated atom, the potential felt by a nearby electron comprises a static component, governed by the ground-state electron-nucleus interaction, and a polarization term arising from the induced multipole moments within the atom. In a liquid environment, this picture becomes more complex. While the central or “focus” atom still polarizes in response to the external electron, neighboring atoms also develop induced dipoles. These surrounding dipoles typically oppose the dipole moment of the focus atom, effectively screening the long-range polarization interaction. This phenomenon results in a net reduction in the attractive polarization component of the electron-atom potential.

To quantify this screening, \cite{Lekner1967} introduced a self-consistent screening function $f(r)$, which accounts for the average influence of the surrounding atomic configuration. The function satisfies the integral equation:
\begin{align}
    f(r) \!=\! 1 \!-\! \pi N\int_0^\infty \!\!\! ds\frac{g(s)}{s^2}\!\!\int_{|r-s|}^{r+s}\!\!dt \frac{\alpha_d(t)f(t)\theta(r,\!s,\!t)}{t^2},
\end{align}
where $r$ is the distance of the electron from the focus atom, $N$ is the number density, $g(s)$ is the pair-correlator, and $\alpha_d$ is the dipole polarizability. The geometric factor $\theta(r,s,t)$ arises from the transformation to bipolar coordinates and is given by:
\begin{align}
    \theta(r,s,t) &= \frac{3}{2}\frac{(s^2 + t^2 -r^2)(s^2 + r^2 - t^2)}{s^2} \nonumber \\
    &\ \ \ \ \ \ \ \ \ \ \ \ \ \ + (r^2 + t^2 -s^2).
\end{align}
In the asymptotic limit of $r\rightarrow \infty$, the screening function approaches the well-known Lorentz screening factor~\citep{Lekner1967}:
\begin{align}
    f_L = \left[1 + \frac{8}{3}\pi N\alpha_d(r\rightarrow\infty)\right]^{-1},
\end{align}
recovering the expected behavior in weakly interacting systems.

\begin{figure}[t]
\begin{center}
\includegraphics[width=8.75cm]{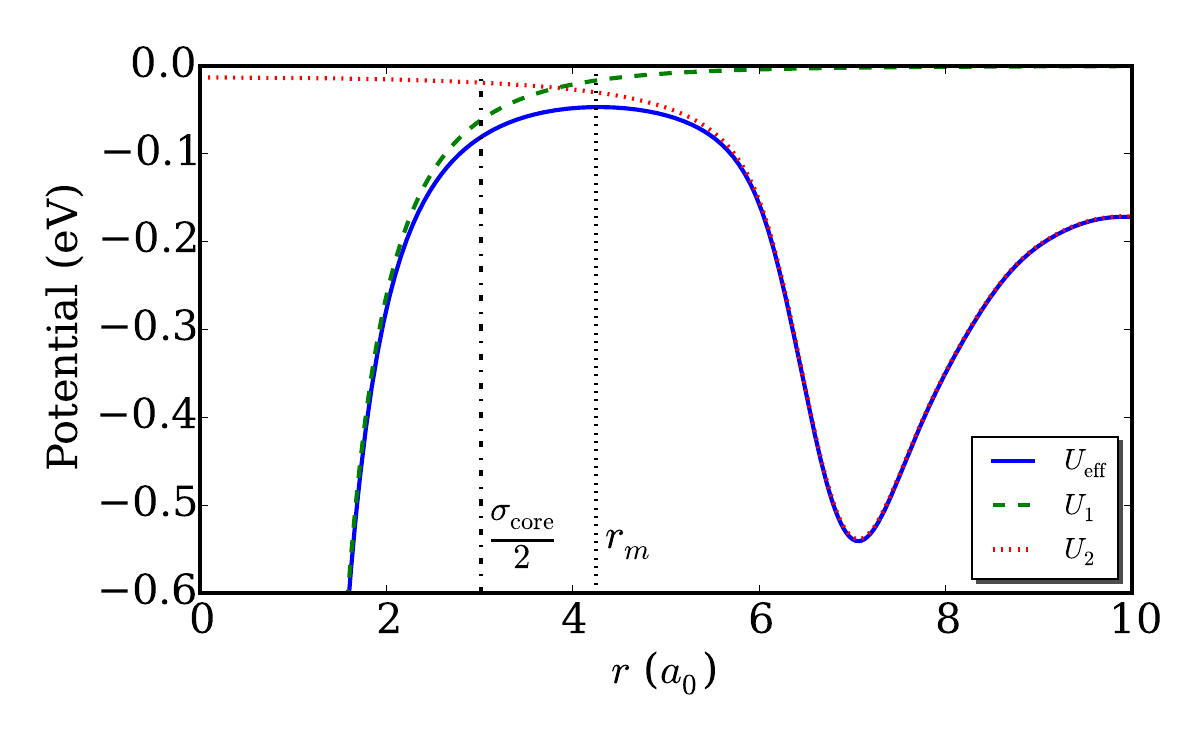}
\end{center}
\caption{Plots of the total effective potential $U_{eff}$ felt by an electron when colliding with one atom in the liquid. Also shown are the components, $U_1$ and $U_2$, which represent the direct potential of the atom and the contribution of the remaining atoms in the bulk, respectively. The dashed vertical lines at $\sigma_{core}/2$ and $r_m$ indicate the hard-core exclusion radius and the proposed collisional sphere, respectively. Note that effects of exchange are not represented in this figure. (Source: Reproduced from~\cite{BoyleEtal2015}, with the permission of AIP Publishing.)}\label{fig:potentials}
\end{figure}

Beyond screening, the surrounding medium also contributes directly to the potential landscape through long-range interactions with neighboring atoms. The total effective scattering potential $U_{eff}$ is thus composed of two terms:
\begin{align}
    U_{eff}(r) = U_1(r) + U_2(r), 
\end{align}
where $U_1(r) = U_{static}(r) + f(r)U_{pol}(r)$ is the screened potential of the focus atom, and $U_2(r)$ accounts for the averaged potential contribution from all other atoms in the bulk. This second term is obtained via:
\begin{align}
    U_2(r) = \frac{2\pi N}{r}\int_0^\infty \!\!dt\ tU_1(t) \int_{|r-t|}^{r+t} \! ds\ g(s).
\end{align}
The effective and potential components for LAr are shown in Figure~\ref{fig:potentials}. These modifications imply that the electron is never truly in a free space environment, i.e., the scattering potential extends throughout the medium. As a result, the use of free-particle plane waves as boundary conditions becomes invalid. Instead, one typically assumes that each atom resides within a finite spherical region, or “scattering cell,” of radius $r_m$, beyond which the potential is treated as approximately constant.

Different criteria have been used to define this matching radius $r_m$. Atrazhev and collaborators adopted the Wigner–Seitz construction, taking  $r_m = (4\pi N/3)^{-1/3}$ as the average interatomic spacing \citep{AtrazhevTimoshkin1996,AtrazhevTimoshkin1998,AtrazhevIakubovPogosov_1995}. In contrast, \cite{Lekner1967} proposed defining $r_m$ as the first turning point of the effective potential, where $\frac{d}{dr}U_{eff} (r_m)=0$. 
At this location, the potential is flat and the electron no longer experiences significant net forces from the surrounding medium. For the LAr case shown in Figure~\ref{fig:potentials}, the value of $r_m \approx 4.3$ a$_0$, which is comparable to Wigner-Seitz diameter of $\approx 4.2$ a$_0$. We also note that $r_m\approx \tfrac{2}{3}\sigma_{\textrm{core}}$ i.e., $r_m$ is larger than half of the minimal interatomic separation $\sigma_{\textrm{core}}$.  

Lekner introduced a shift $V_0$ to the effective total potential such that~$U_{eff}(r_m)~+~V_0~=~0$, ensuring that the potential vanishes beyond the matching radius, simplifying the application of asymptotic boundary conditions in partial wave expansions. An alternative approach, used in more recent work~\citep{BoyleEtal2015,BoyleEtal2016,WhiteEtal2018}, involves calculating the scattering phase shifts directly at $r_m$. This avoids artificial shifts in the potential and more accurately captures the available energy states within the liquid, especially near the conduction band minimum.

The quantity $V_0$, referred to as the background energy or conduction band minimum of the liquid, plays a central role in defining the energy scale for electron transport. Analogous to the band structure in solids, it sets the zero-point energy for delocalized electrons in the liquid. Numerous methods have been developed to estimate $V_0$, both theoretically~\citep{SpringettCohenJortner1967,Hernandez1991,NazinShikin_2005,IakubovKhrapak1982,Nieminen_1980} and experimentally~\citep{EvansFindley2010,EvansKyrnskiStreeterFindley2015}.

These modifications to the scattering potential are essential for accurate modeling of electron transport in dense media. They determine not only the magnitude and energy dependence of scattering cross-sections but also the reference energy scale against which all transport properties, such as mobility, diffusion, and drift energy, are defined.

\subsubsection{Example transport calculations: electrons in liquid Ar}\label{sub:liqAr}

To illustrate the practical application of the \textit{ab initio} framework described in the previous sections, we now consider electron transport in LAr~\citep{BoyleEtal2015}. The same methodology, with minor adaptations, has been successfully extended to other systems, including electron transport in LXe~\citep{BoyleEtal2016}, liquid krypton~\citep{WhiteEtal2018}, and positron transport in liquid helium~\citep{Cocks2020}.

\begin{figure}[t]
\begin{center}
\includegraphics[width=8.5cm]{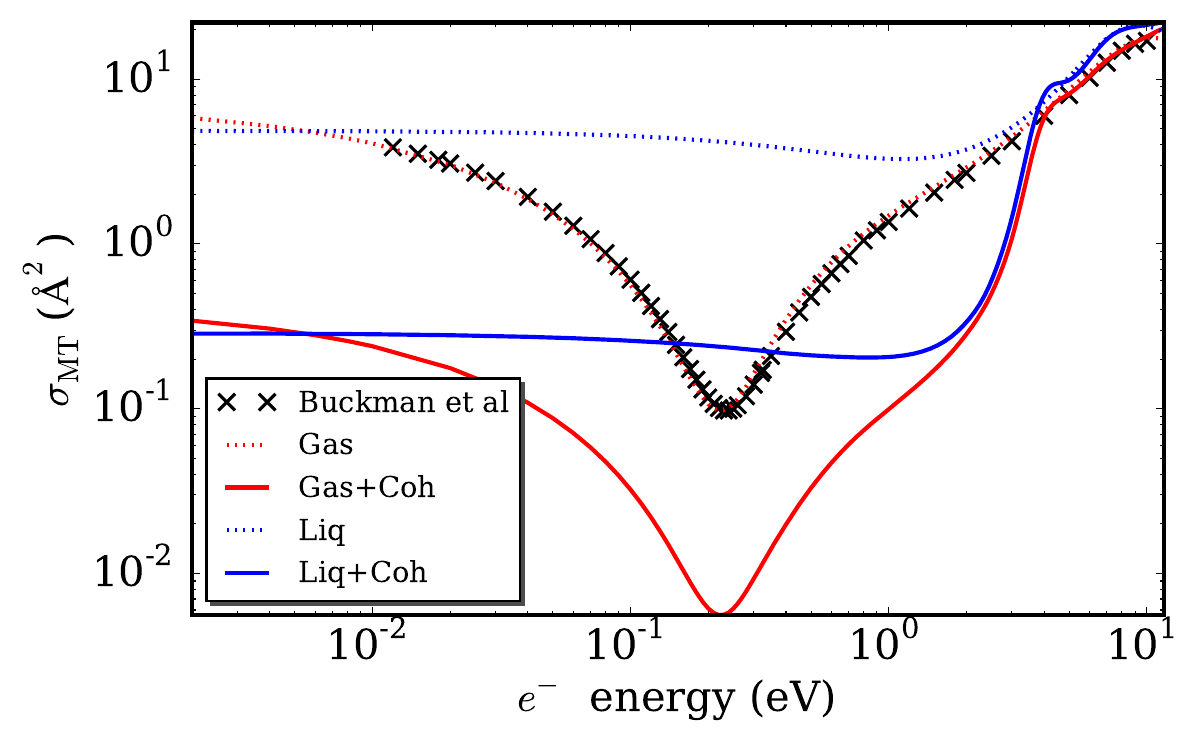}
\end{center}
\caption{The momentum-transfer cross-sections in the gas-phase (Gas) and liquid-phase (Liq) and their modifications when coherent scattering effects are included (+Coh). The recommended transfer cross-section of \cite{BuckmanEtal2000} for a dilute gas is a combination of experimental measurements and theoretical calculations. (Source: Reproduced from~\cite{BoyleEtal2015}, with the permission of AIP Publishing.)}\label{fig:MTXsects}
\end{figure}

At low reduced electric field strengths (typically below a few Townsends), elastic scattering dominates, and the modifications to the scattering potential and coherent effects described earlier become particularly significant. Figure~\ref{fig:MTXsects} presents the momentum-transfer cross-sections for electron-argon scattering in both the gas and liquid phases, elucidating the separate roles of coherence and the potential modifications. For validation, the dilute gas-phase results are compared with the benchmark data of~\cite{BuckmanEtal2000}, which represent a synthesis of high-quality experimental measurements and theoretical predictions. The agreement confirms the accuracy of the underlying electron-argon interaction potential and the reliability of the solution technique employed.

Notably, the influence of screening and the liquid environment significantly alters the cross-section profile. In the gas phase, a pronounced Ramsauer minimum is evident, reflecting a reduced probability of momentum-transfer at specific electron energies. However, in the liquid phase, this feature is suppressed due to the screening of the polarization potential by surrounding atoms. As a result, the cross-section becomes nearly energy-independent at low incident energies, a hallmark of scattering in dense fluids. The inclusion of coherent scattering further reduces the momentum-transfer cross-section in the low-energy regime. At higher energies, where the electron’s de Broglie wavelength is much shorter than the interatomic spacing, the effects of coherence and screening diminish, and the liquid-phase cross-section asymptotically approaches its gas-phase counterpart.

\begin{figure}[t]
\begin{center}
\includegraphics[width=8.5cm]{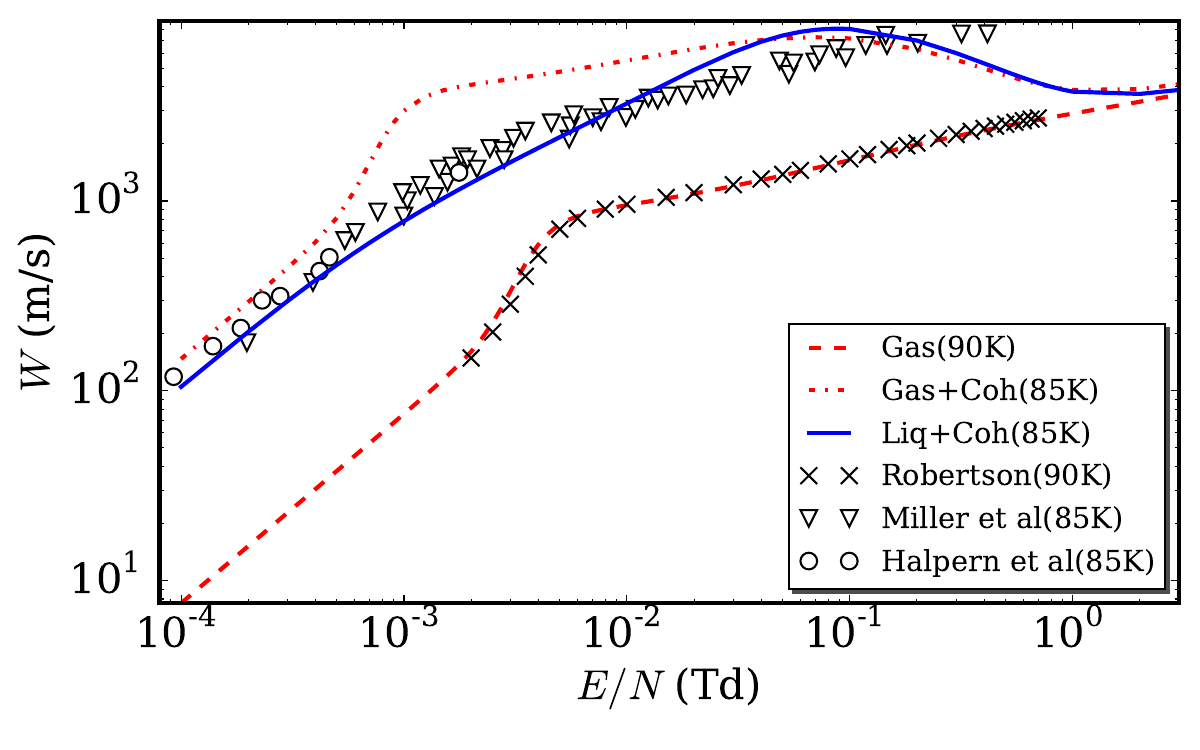}
\\
\includegraphics[width=8.5cm]{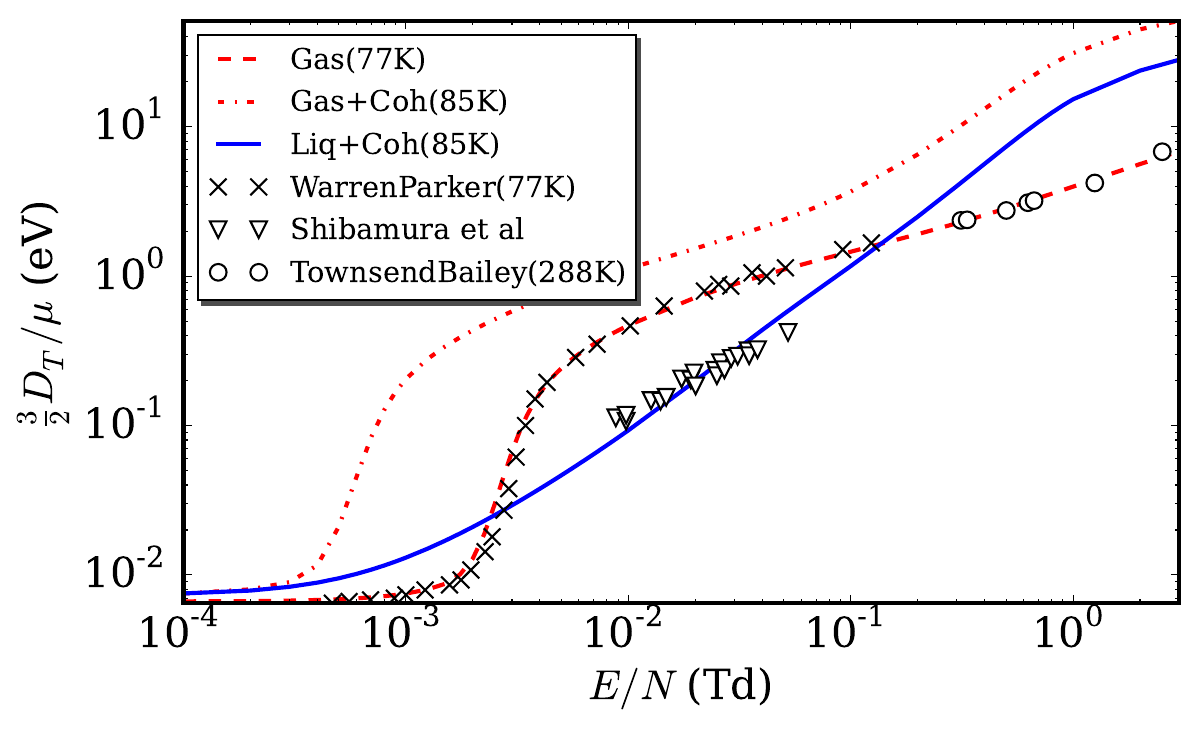}
\end{center}
\caption{Comparison of the measured drift velocities $W$ (top) and
characteristic energies $\tfrac{3}{2} D_T/\mu$ (bottom) in gaseous and liquid argon, with those calculated from the various approximations to the cross-sections. Experimental data—Ar: \cite{Robertson1977} at 90 K;
\cite{Miller1968} at 85 K; \cite{Halpern1967} at 85 K; \cite{WarrenParker_1962} at 77 K; \cite{TownBail22} at 288 K; \cite{Shibamura1979} at an unmeasured liquid temperature. The various approximations used are: gas phase only cross-sections (Gas), gas phase cross-sections with coherent scattering (Gas+Coh), and liquid-phase cross-sections with coherent scattering effects (Liq+Coh). The results have been calculated using the full differential cross-section and results are converged multi-term values. (Source: Reproduced from~\cite{BoyleEtal2015}, with the permission of AIP Publishing.)}\label{fig:TransCoLiq}
\end{figure}

Figure~\ref{fig:TransCoLiq} shows the calculated drift velocity and characteristic energy for electrons in LAr as functions of the reduced electric field. These transport properties were obtained using the cross-sections derived from the \textit{ab initio} framework and solved via the multi-term BE formalism described in Section~\ref{sub:Boltz}. For comparison, selected experimental measurements are included to validate the model. The results clearly demonstrate that applying gas-phase cross-sections scaled by density is insufficient to capture the correct transport behavior in liquids. While the gas-phase potential accurately describes transport in dilute conditions, it fails to account for the dense-medium effects that dominate in the liquid state.

Importantly, the separate impacts of coherent scattering and the potential modifications are highlighted. Coherent scattering alone is sufficient to bring the drift velocity into the correct order of magnitude. However, it overestimate the characteristic energy, further increasing the discrepancy with experimental liquid data. When combined with the potential modifications, the transport properties closely match the experimental observations, underscoring the necessity of accounting for both effects simultaneously to achieve quantitative agreement.

These results highlight the strength of the \textit{ab initio} modeling approach in reproducing transport behavior in liquids, without the need for empirical tuning. They also emphasize the inadequacy of using gas-phase data alone in high-density regimes, and the importance of including medium-induced modifications to the scattering process in any predictive transport framework.

\subsubsection{Inelastic scattering}\label{sub:inelastics}

While elastic scattering dominates low-energy electron transport in noble liquids, inelastic processes become increasingly relevant near and above excitation thresholds. At temperatures just below the melting point, the reflection spectrum of LXe closely resembles that of solid xenon, though the exciton lines in the liquid are shifted to lower energies and broadened due to increased structural disorder~\citep{Beaglehole1965}.

These shifts have been traced to changes in the band gap, i.e., the energy difference between the uppermost valence band and the lowest conduction band, which decreases with increasing temperature and density~\citep{Steinberger1973}. This behavior is attributed to the movement of band edges as the lattice dilates. Notably, the position of excitonic features in LXe correlates linearly with the fluid’s density, suggesting that interatomic distances, rather than long-range order, primarily determine the excitonic energy levels.

Spectroscopic studies have revealed several intermediate excitons in LXe, with characteristics of both Frenkel and Wannier excitons for $n=1$. Prominent excitonic bands appear at approximately 8.3~eV, 9.5~eV, and 10.3~eV \citep{Beaglehole1965,Asaf1971,Laporte1977,Laporte1980,ReiningerEtal_1983}. The first two excitonic bands have parentage in the atomic transitions 6s(3/2)$^{0}_{1}$ (5p$^5$6s [K=3/2]) at about 8.44~eV, and 6s'(1/2)$^{0}_{1}$ (5p$^5$6s [K=1/2]) at about 9.57~eV, respectively The third peak arises from the two neighboring bands which correspond to 5d(3/2)$^{0}_{1}$ (5p$^5$5d [K=3/2]) at about 10.40~eV and 7s(3/2)$^{0}_{1}$. For instance the intermediate exciton at 9.5~eV may be identified both as the $^{1}$P$_{1}$ Frenkel exciton and as the $n=1$ member of the [$\Gamma(1/2)$] series in the Wannier notation \citep{Laporte1977,Laporte1980}. Similarly, the intermediate exciton at 8.3~eV is associated both with the $^{3}$P$_{1}$ exciton in the Frenkel scheme, and as the $n = 1$ member of the [$\Gamma(3/2)$] Wannier exciton series.

Higher-order excitons, such as the $n = 2$ member of the [$\Gamma(3/2)$] series, have also been observed, albeit with much weaker spectral features~\citep{Asaf1971,Asaf1974,Laporte1980}. In addition to excitonic lines, broadened and shifted atomic excitation features have been identified in the reflection spectra, particularly near 8.3~eV, where they overlap with excitonic transitions~\citep{Asaf1971,Laporte1977,Laporte1980}.

The emergence of excitonic features in the reflection spectrum is strongly density-dependent. Studies show that excitonic lines begin to appear at densities significantly below the liquid state, with no abrupt transition from the ``dense gas" to ``liquid" regime, even along continuous thermodynamic paths where the entire system is a single-phase one~\citep{Laporte1977,Laporte1980}. Exciton formation requires local clusters of atoms large enough to contain an exciton within its characteristic radius. Where this condition is not met, only broadened atomic transitions are observed. Both the excitonic lines and the perturbed atomic
excitations exist in LXe, and they both contribute to energy losses of quasifree electrons in the conduction band, complicating first-principles model development.

Photoconductivity measurements provide a complementary insight. The conduction band minimum in LXe, corresponding to the $\Gamma(3/2)$ gap, has been determined to lie at 9.22~eV \citep{Asaf1974}, in agreement with estimates based on density changes upon melting. A dip in the photoconductivity spectrum near~9.45 eV coincides with the $n=1$ member of [$\Gamma(1/2)$] exciton, suggesting competition between optical excitation and free-carrier generation. At higher energies, up to about 10.0~eV no further structure in the photoconductivity spectra has been observed~\citep{Asaf1974}. 

The evolution of the photoconductivity spectra of fluid xenon with increasing density from 10$^{21}$ atoms/cm$^3$ up to the triple point density has been thoroughly investigated by \cite{ReiningerEtal_1983}. It has been found that the photoconductivity threshold decreases with increasing density. Its extrapolation to zero density yields 11.10~eV which is the difference between the energy minimum of the Xe$_{2}^{+}$ molecular ion and the ground-state energy of the xenon atom. It has been observed that there is a decrease in photocurrent at about 10.32~eV, where a reflectivity maximum has been observed \cite{ReiningerEtal_1983}. While for the interband transitions the electrons and holes become quasifree, this discrete bound state likely has alternative decay channels like luminescence, which reduces the measured photocurrent at the line. At higher liquid densities and in the solid phase, the discrete 9.45~eV reflection peak appears as a pronounced dip in the photoconductivity spectrum.

Despite the wealth of spectroscopic data, theoretical treatments of inelastic scattering in noble liquids remain limited. In early transport studies of LXe and LAr~\citep{Garland2018,SimonovicEtal_2019}, inelastic collisions were treated approximately, due to the lack of reliable cross-section data for exciton formation, perturbed atomic excitations, and interband transitions in the liquid phase. Crucially, the fraction of clusters that support exciton formation versus those contributing only perturbed atomic transitions remains unknown, especially for optically forbidden transitions that still contribute to inelastic energy loss for conduction-band electrons.

To address this, \cite{Garland2018} and \cite{SimonovicEtal_2019} approximated the interband excitation cross-sections in LXe and LAr using gas-phase electron impact ionization cross-sections, adjusted to match the known liquid band gaps. Similarly, excitonic and atomic excitation cross-sections were approximated using those of isolated atoms. Wannier excitons with $n>1$ were omitted, as their parentage lies beyond atomic excited states, and only weak spectral signatures (e.g., the $n=2$ exciton at 9.0 eV) have been observed~\citep{Asaf1971}.

For the case of LXe, \cite{SimonovicEtal_2019} systematically explored four scenarios for including inelastic excitations in their transport model, using the cross-section set from ~\cite{Hayashi2003Xe}:
\begin{enumerate}
    \item No excitations included, i.e., assuming purely elastic and ionizing collisions.
    \item Only excitations below the band gap included, consistent with the approach in \cite{Garland2018} for LAr.
    \item First four excitations from the Hayashi cross-section set included, to account for the optically
allowed excitation which leads to the pronounced dip in the
photoconductivity spectra of LXe at about 9.45 eV as well as one
optically forbidden excitation which has a slightly lower threshold. 
    \item All excitations in the Hayashi set included, extending coverage to features up to 10.3 eV.
\end{enumerate} 

\begin{figure}[b]
	\includegraphics[scale=0.32]{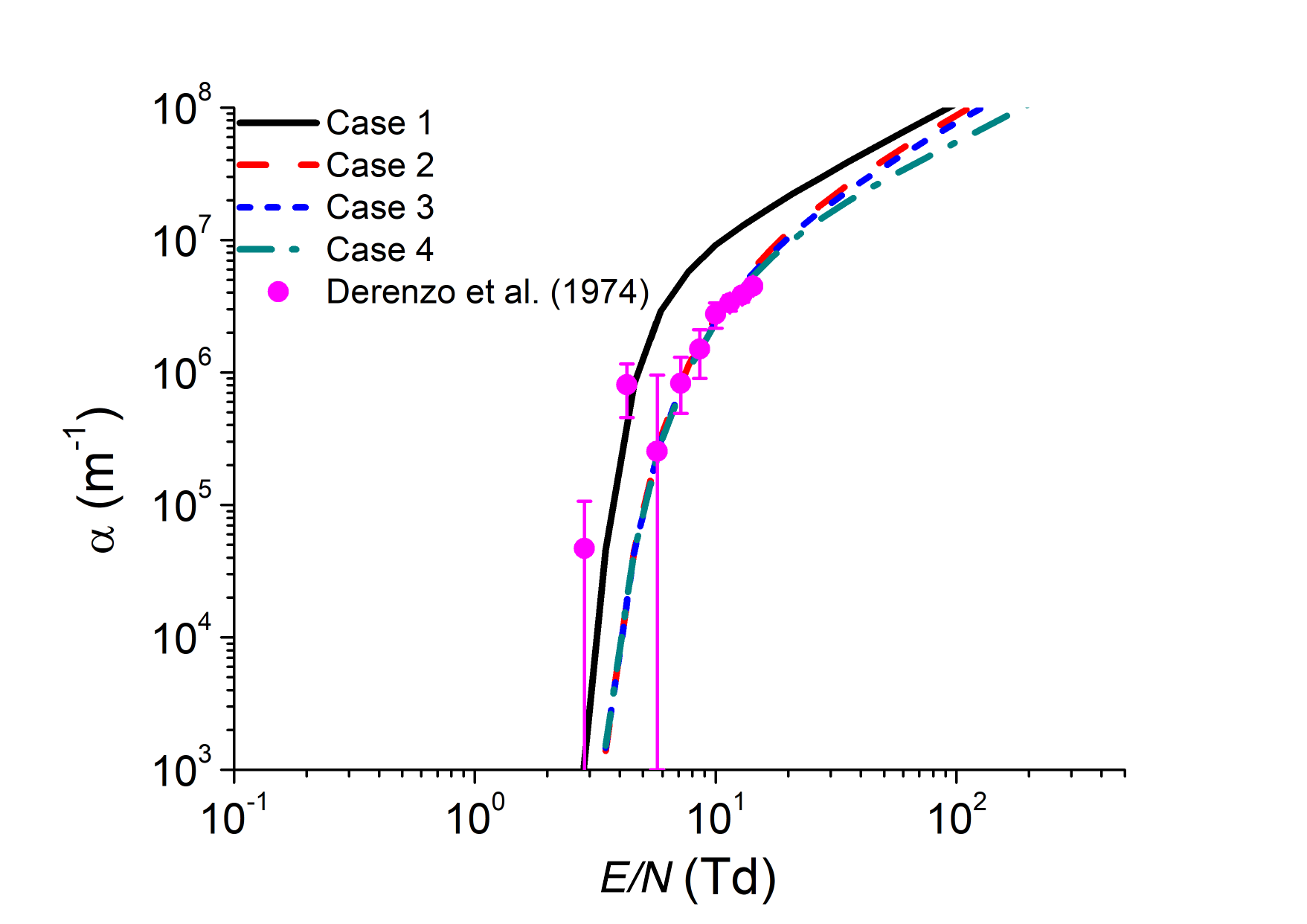}
	\caption{Comparison of the ionization coefficient $\alpha$ for electrons in LXe calculated by \cite{SimonovicEtal_2019} for four different treatments of inelastic collisions (see text for details) with the experimental measurements of ~\cite{Derentzo1974}.}
	\label{fig:LXeComparison} 
\end{figure}

For each scenario, the ionization coefficient in LXe was calculated and compared to experimental data from \cite{Derentzo1974}, and is shown in Figure~\ref{fig:LXeComparison}. Interestingly, the first two experimental points matched best with the first scenario (no excitation), while the remaining data aligned closely with the fourth scenario (all excitations included).

However, it is important to note that the experimental ionization data are limited to relatively low reduced electric fields, well below those used in advanced detectors. Thus, while the fourth case shows good agreement at low fields, it is uncertain whether these approximations remain valid at higher fields. At elevated energies, the detailed structure of the conduction band and the exact inelastic scattering dynamics may play a more prominent role, requiring a more rigorous theoretical treatment.

\subsection{Electron self-trapping}\label{sub:trapping}

In liquid-phase systems, electron transport is not always governed solely by scattering and diffusion. Under certain conditions, excess electrons may become transiently localized in the medium due to spontaneous density fluctuations. This phenomenon, known as electron self-trapping, results in the formation of localized states, such as bubbles or solvated electron configurations, that significantly alter the transport dynamics~\citep{Nieminen_1980,Maris2003, Rumbach2015,Abel2013,Gopalakrishnan2016}.

The driving mechanism behind self-trapping lies in the nature of the electron–atom interaction in noble liquids. These interactions are typically dominated by short-range repulsion, resulting in a positive background potential $V_0(N)>0$, where $N$ is the local atomic density. An electron can lower its free energy by migrating into regions of reduced density, where the value of $V_0$ is lower. If the density depression is large enough, the energetic gain from a reduced potential energy can outweigh both the kinetic energy cost of localization and the mechanical work needed to expand the local cavity \citep{Borghesani1992a,Hernandez1991}. In such cases, a metastable localized state may be formed. These states are weakly bound and have been observed in both theoretical studies and experimental measurements of atomic liquids~\citep{Borghesani2006}.

\cite{Borghesani2006} proposed a phenomenological model for mobility in systems exhibiting self-trapping, treating the total electron mobility as a weighted average of contributions from two populations: free (delocalized) electrons and localized electrons trapped in density fluctuations. The mobility of the free electrons is computed using a heuristic kinetic model, while the contribution from localized electrons is modeled using hydrodynamic descriptions of bubble transport. Although this approach qualitatively reproduces observed trends, it struggled to match absolute mobility values, primarily because the assumption of a single, well-defined localized state is an oversimplification. In reality, liquid systems support a broad distribution of fluctuation sizes and configurations, each contributing differently to the transport properties.

To address these limitations, more sophisticated approaches have been developed. \cite{CocksWhite_2016} performed \textit{ab initio} calculations of the probability that an electron scatters into a low-density region and becomes trapped. Their model accounts for both the formation and energetic stability of these localized states and provides a physically grounded framework for studying electron localization in liquids.

Further progress was made by Stokes and collaborators~\citep{Stokes_etal_2016}, who introduced a generalized BE capable of describing the interplay between localized and delocalized electron populations. Their framework incorporates scattering into and out of density fluctuations, allowing for both trapping and detrapping processes to be treated self-consistently. This hybrid transport model provides a more complete picture of electron dynamics in systems where localization phenomena are non-negligible.

Overall, electron self-trapping represents a key departure from conventional gas-phase transport behavior and introduces rich physics into the study of dense, disordered systems. Accurate treatment of this effect is essential for modeling electron mobility in noble liquids, particularly under low-field or low-temperature conditions where fluctuations are more prominent and localization becomes more probable.

\section{Applications and challenges for liquid detectors}\label{sec:applications}

While much progress has been made in understanding the fundamental physics of electron transport in noble liquids, practical applications in detector design introduce additional complexities. In this section, we examine selected challenges and applications relevant to noble-liquid detectors, with a focus on scintillation and luminescence, doped and mixed-liquid systems, and charge transport across gas-liquid interfaces.

\subsection{Scintillation and Luminescence}\label{sub:scintilation}

Studies of ionisation, scintillation, and luminescence in noble gases and liquids, particularly in LAr and LXe, play a vital role in the development of modern detector technologies~\citep{Gonzalez2018,Bonivento2024}. When a charged particle traverses a noble liquid, energy dissipation occurs via ionization, excitation, and the generation of sub-excitation electrons. The average energy loss per ionization event slightly exceeds the ionization potential or the energy for the interband transition, as it encompasses multiple ionisation events. Notably, the average ionization energy is lower in the liquid phase compared to the gas, implying more efficient ionization. Among noble liquids, LXe exhibits a lower energy cost per ion pair than LAr, resulting in a higher ionization yield. The liberated electrons then drift and diffuse under an applied electric field, enabling both calorimetric and spatial reconstruction in TPC detectors~\citep{AprileBaudis2010}.

Scintillation, the emission of light following atomic or molecular excitation, plays a central role in signal formation. More broadly, luminescence includes all photon emission processes (scintillation among them) arising from energy deposition and subsequent relaxation of excited states. In the context of liquid noble gases, these terms are often used interchangeably, and we refer to all prompt light emission as scintillation.

In LAr and LXe, scintillation is predominantly driven by the formation and subsequent de-excitation of dimers, also known as excimers, i.e., transient, bound states formed when an excited atom (Ar* or Xe*) interacts with ground-state atoms. These excimers exist in singlet and triplet configurations, which de-excite by emitting vacuum ultraviolet (VUV) photons: approximately 128~nm in LAr and 178~nm in LXe. Time-resolved spectroscopy confirms this mechanism, showing distinct fast and slow scintillation components that correspond to singlet and triplet state lifetimes ~\citep{Kubota1978,Hitachi1983,Segret02021}. Further experimental evidence includes the dependence of scintillation yield and timing on factors such as electric field intensity, particle type, and energy
deposition \citep{Doke2002,Doke2006}. These observations are consistent with theoretical models that describe the formation and decay of excimers.

Direct de-excitation of excited atoms also contributes to scintillation, though to a lesser extent. For example, in LXe, isolated Xe* atoms can emit photons upon returning to the ground state. However, in the dense liquid environment, this pathway is suppressed in favor of excimer (Xe$_2$*) formation via collisions between excited and ground-state xenon atoms. Although the direct de-excitation of excited LXe atoms plays a relatively minor role in the scintillation process compared to the excimer mechanism, it does contribute to the overall luminescence.

\begin{figure}[t]
	\includegraphics[scale=0.32]{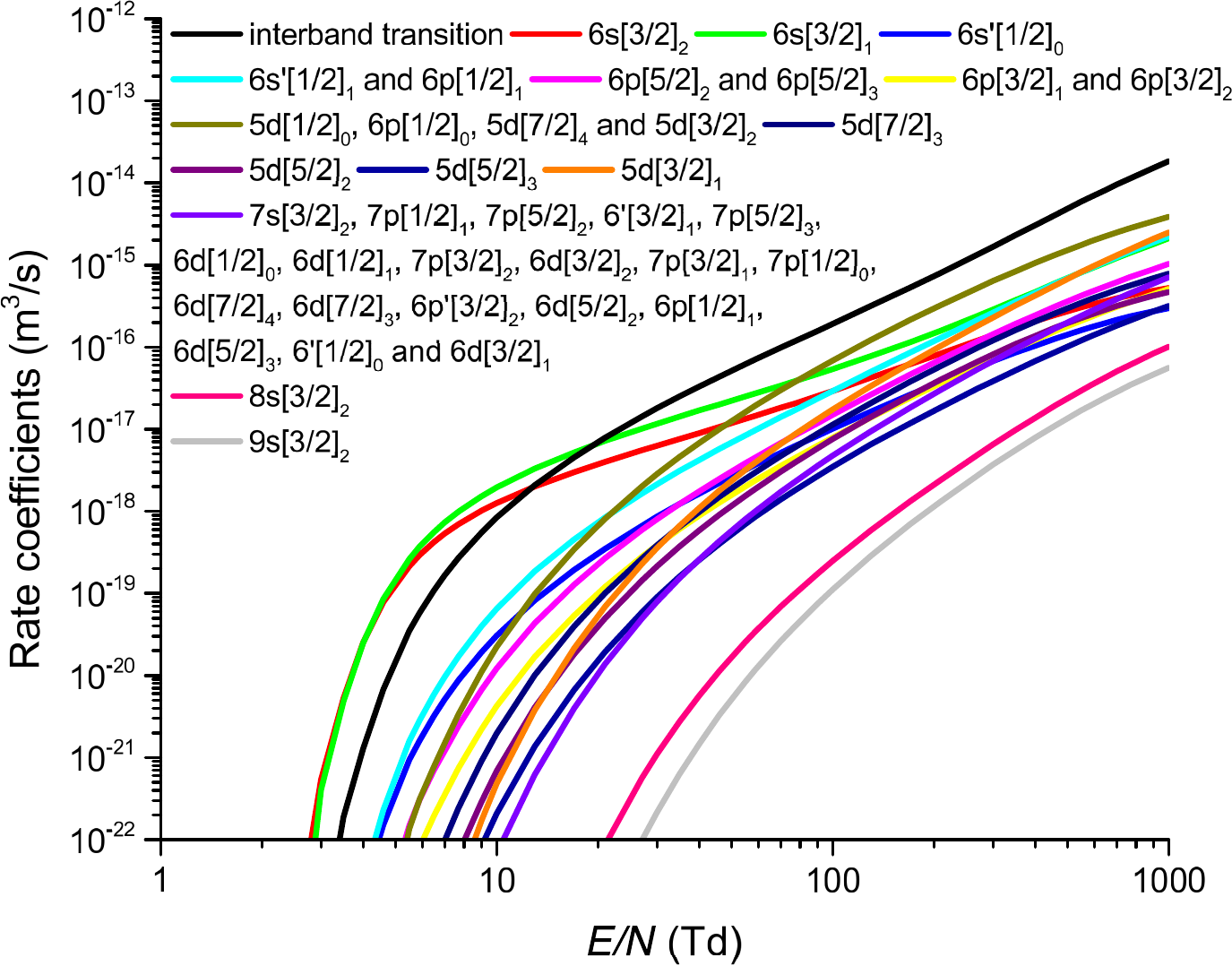}
	\caption{Rate coefficients for interband transition and electronic excitations in LXe as a function of $E/N$. See text for details.}
	\label{fig:LXeRates}
\end{figure}

To model these processes quantitatively, accurate rate coefficients for inelastic collisions, such as atomic excitations and interband transitions, are required. Figure~\ref{fig:LXeRates} illustrates rate coefficients computed using MC simulations and electron scattering cross-sections from \cite{SimonovicEtal_2019}. At low reduced fields, $E/N$, excitation and interband transitions are negligible. However, as $E/N$ increases beyond 20 Td, the interband transition rate becomes dominant. The computed rate coefficients are smooth and featureless, reflecting the threshold-like onset of inelastic processes as electrons gain sufficient energy. These results are essential for refining empirical simulation tools such as NEST ~\citep{Szydagis2011,Szydagis2021,szydagis2025review}. In addition, the rate coefficients for various inelastic processes are required as input for the fluid modeling of discharges in LXe.

Doping is a powerful technique for enhancing detector performance. For example, introducing trace amounts of LXe into LAr shifts the emission spectrum from 128 nm to ~175–178 nm. This shift improves photon detection efficiency by reducing Rayleigh scattering and matching the spectral sensitivity of common photodetectors. Even at concentrations as low as a few parts per million, xenon doping can significantly increase scintillation yield by enabling energy transfer from LAr to LXe excimers~\citep{Segret02021,Fields2023,DUNE2024}. A first step towards accurate \textit{ab initio} modeling of doped and simple liquid mixtures is detailed in Section~\ref{sub:mixtures}.

In contrast, impurities such as nitrogen, oxygen, and water vapor degrade detector performance by quenching scintillation and capturing free electrons. These species either absorb VUV photons or serve as electron scavengers, reducing both light output and charge collection. Oxygen, in particular, has a high electron affinity and is highly disruptive to charge readout~\citep{Acciarri2010,Barrelet2002}. Modeling studies have been conducted to understand impurity dynamics in LAr detectors, including their sources, transport, and removal mechanisms. To mitigate these effects, sophisticated purification systems and impurity monitoring tools, such as ultra-sensitive trace gas sensors developed for the DARKSIDE experiment, have been employed to maintain purity at sub-part-per-billion levels~\citep{Mount2012}.

Overall, the interplay between excitation, ionization, and scintillation mechanisms, along with the impact of doping and impurity control, defines the performance of noble-liquid detectors. Accurate modeling of these processes, supported by both experimental data and advanced simulations, remains critical for optimizing detector response and enabling precision measurements in neutrino and dark matter experiments.

\subsection{Electron transport in doped and mixture systems}\label{sub:mixtures}   

The ECFA DRD roadmap~\cite{EFCA2021} highlighted the potential benefits of using liquid mixtures and dopants as a detector medium. For instance, doping LAr with small concentrations of LXe has been shown to enhance charge amplification by over two orders of magnitude~\citep{Kim_etal2002}, offering a promising route for improved detector performance~\citep{Segret02021,Fields2023,DUNE2024}. As discussed in Section~\ref{sub:scintilation}, even trace amounts of LXe in LAr reducing Rayleigh scattering, matches the spectral sensitivity of common photodetectors, and enables energy transfer from LAr to LXe excimers.

In Section~\ref{sub:liquids}, we introduced the formalism for coherent elastic scattering in pure single-component liquids. Extending this framework to mixtures involves generalizing the collision operator to account for multiple species. In the case of a binary mixture (e.g., a doped LAr–LXe system), the coherent elastic collision operator generalizes to~\citep{Boyle_2024}:
\begin{align}
    &J_0 (\Phi_0) = \frac{m_e}{m_av^2}\left[v\nu^{a}_1(v)\left(v\Phi_l + \frac{k_bT}{m_e}\frac{d}{dv}\Phi_0\right)\right] \nonumber \\ 
    &\ \ \ \ \ \ + \frac{m_e}{m_bv^2}\left[v\nu^{b}_1(v)\left(v\Phi_l + \frac{k_bT}{m_e}\frac{d}{dv}\Phi_l\right)\right]\!, \label{eq:J0mix} \\
    &J_l (\Phi_l) = \tilde{\nu}_l(v) \Phi_l,\ \ \ \ \text{for } l \geq 1, \label{eq:Jlmix} 
\end{align} 
where $\nu^{a}_l(v)$ and $\nu^{b}_l(v)$ are the usual Legendre components of the binary collision frequencies for each species:
\begin{align}
    \frac{\nu^{a,b}_l(v)}{x_{a,b}N} &= 2\pi v\!\!\int_0^{2\pi}\!\!\sigma_{a,b}(v,\theta)[1-P_l(\cos\theta)]\sin\theta d\theta.
\end{align}
The structure-modified collision frequency $\tilde{\nu}_l(v)$ is now calculated from the combined differential cross-section:
\begin{align}
    \Sigma(v,\theta)
    &=  \sigma_a(v,\theta)S_{aa}\!\left(K\right) + \sigma_b(v,\theta) S_{bb}\!\left(K\right) \nonumber\\
    &\ \ \ \ \ \ \ \ \ + 2 \sqrt{\sigma_a(v,\theta)\sigma_b(v,\theta)} S_{ba}\!\left(K\right)\!. \label{eq:DCSmix}
\end{align}
where $K = \frac{2m_ev}{\hbar}\sin{\frac{\theta}{2}}$, and  $S_{aa}$, $S_{bb}$ and $S_{ab}$ are the partial static structure factors describing intra- and inter-species correlations (see~\cite{Boyle_2024,HansenMcDonald1973} for details). In the dilute limit where $K$ is large, the structure factors approach their respective mole fractions ($S_{aa}(K) \rightarrow x_a, S_{bb}(K) \rightarrow x_b$ and $S_{ab}(K) \rightarrow 0$) and the usual binary scattering results are recovered, as required. The formalism also correctly reduces to the single-species case when $\sigma_b \rightarrow \sigma_a$, i.e., $S_{aa} + 2S_{ab} + S_{bb} \rightarrow S_a$, where $S_a$ represents the structure factor of the pure fluid.

\begin{figure}[t]
\begin{center}
\includegraphics[width=8.5cm]{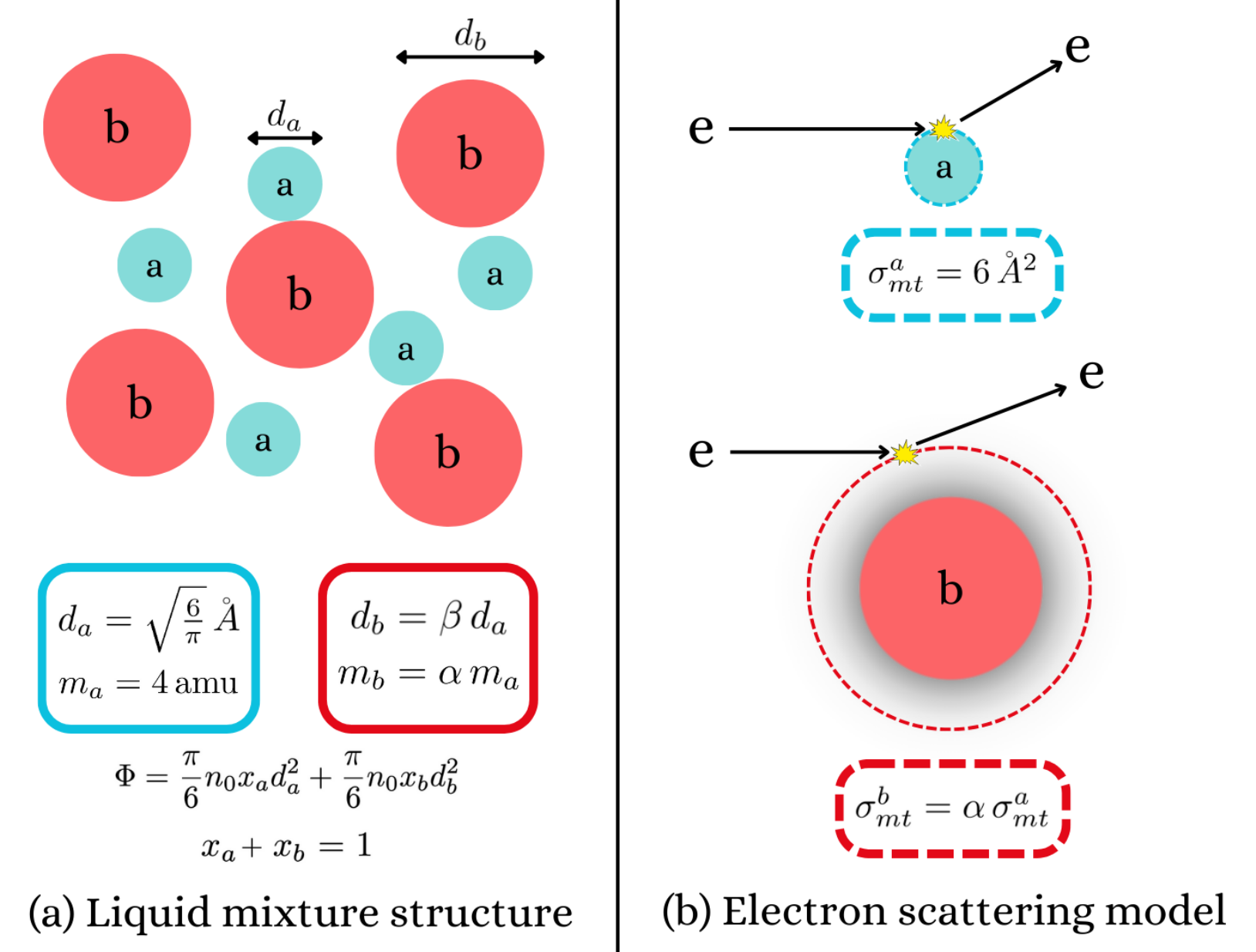}%
\end{center}
\caption{Benchmark binary system of cold ($T=0$ K) hard-spheres. \textbf{a)} Parameters describing the liquid mixture structure: $d_a$ ($d_b$), $m_a$ ($m_b$) and $x_a$ ($x_a$) are the hard-sphere diameter, mass and density fraction of species $a$ (species $b$), respectively. $\Phi$ is the total packing factor, $N$ is the total number density. \textbf{b)} Parameters describing the electron scattering: $\sigma_{mt}^a$ ($\sigma_{mt}^b$) is the momentum-transfer cross-section for electron scattering from species $a$ (species $b$), and $\sigma_0^a-\sigma_l^a = \sigma_{mt}^a$ ($\sigma_0^b-\sigma_l^b = \sigma_{mt}^b$) is constant for all $l$. (Source: Reproduced from~\cite{Boyle_2024}, with the permission of IOP Publishing.)}\label{fig:MixtureDiagram}
\end{figure}

The static structure factors and pair distribution functions required in these calculations can be obtained via molecular dynamics or MC simulations~\citep{GunsterenBeredsen90,Lindgard1994,WedbergEtal2011}. However, for hard-sphere liquids, an analytic solution is available from the Percus–Yevick approximation~\citep{PercusYevick1958,Lebowitz1964,Hiroike1969}. For binary mixtures, the structure is fully specified by ratio of hard-sphere diameters of the two species $\beta = \frac{d_b}{d_a}$, the diameter and density fraction of species $a$ ($d_a,x_a$), and the total packing fraction $\Phi$:
\begin{align}
    \Phi = \frac{\pi}{6}N x_ad_a^2 + \frac{\pi}{6} N x_bd_b^2. \label{eq:phi}
\end{align}
where  $x_b = 1 - x_a$, and $N$ is the total number density. These parameters define the extent of spatial correlation within the mixture and strongly influence electron transport. We now consider the benchmark model introduced in \cite{Boyle_2024}, shown in Figure~\ref{fig:MixtureDiagram}.

\begin{figure}[t]
\begin{center}
\includegraphics[width=9cm]{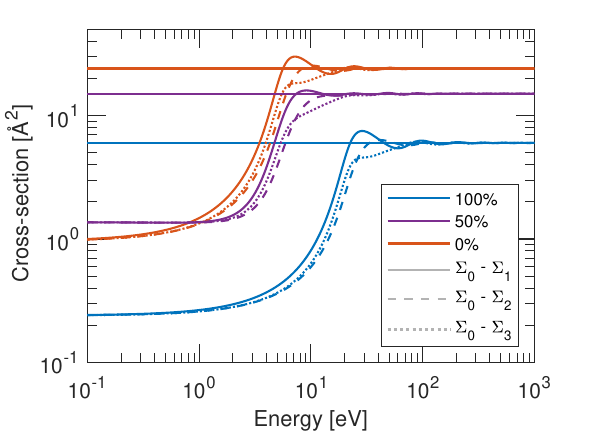}
\end{center}
\caption{Scattering cross-section variation with electron energy for the binary hard-sphere mixture given in Figure~\ref{fig:MixtureDiagram}. Partial cross-sections $\Sigma_0 - \Sigma_l$ vs energy are given for varying percentages of species $a$, and with parameters $\Phi = 0.4$, $\alpha = \frac{\sigma_b}{\sigma_a} = 4$, $\beta = \frac{d_b}{d_a} = 2$. Horizontal lines represent the (constant) binary-scattering limit cross-sections. (Source: Reproduced from~\cite{Boyle_2024}, with the permission of IOP Publishing.)}\label{fig:MixtureXsect}
\end{figure}

Figure~\ref{fig:MixtureXsect} illustrates the structure-modified collision components  $\Sigma_0 - \Sigma_l$ for three different compositions: pure species $a$ ($x_a = 1$), pure species $b$ ($x_a = 0$), and a 50:50 mixture ($x_a = 0.5$). Similar to the pure liquid case, the structure-modified profiles exhibit oscillations around the constant binary scattering value and show significant suppression of momentum-transfer at low energies. Interestingly, at energies near 0.1 eV, the 50:50 mixture profile exceeds those of both pure systems, indicating non-linear mixing behavior in the cross-sections.

\begin{figure}[h!]
\begin{center}
\includegraphics[width=8.0cm]{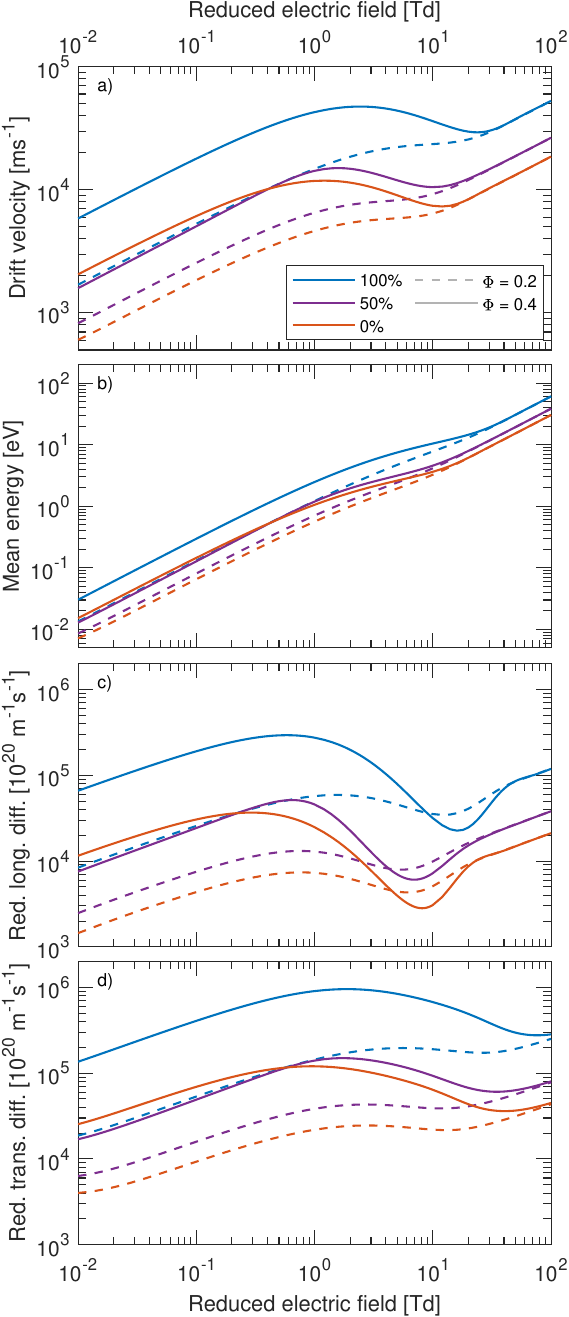}
\end{center}
\caption{Transport properties 
for varying  mixture percentage of species $a$. The parameters of species $a$ and $b$ are set according to the model given in Figure~\ref{fig:MixtureDiagram}, with $\alpha = \frac{\sigma_b}{\sigma_a}=4$ and $\beta = \frac{d_b}{d_a}=2$. \\ \textbf{a)} Drift velocity, \textbf{b)} Mean energy, \textbf{c)} Reduced longitudinal diffusion coefficient, \textbf{d)} Reduced transverse diffusion coefficient. (Source: Reproduced from~\cite{Boyle_2024}, with the permission of IOP Publishing.)}\label{fig:MixtureTransport}
\end{figure}

These effects are directly reflected in the calculated transport coefficients. Figure~\ref{fig:MixtureTransport} shows the variation of drift velocity $W$, mean energy $\epsilon$, and reduced diffusion coefficients $ND_L$ and $ND_T$, as functions of the reduced electric field. The results are shown for two packing fractions ($\Phi = 0.2$ and $\Phi = 0.4$) with $\beta = 2$ and $\alpha = \beta^2 = 4$. Large variations spanning orders of magnitude are observed across different mixture compositions. Notably, at the higher packing density ($\Phi = 0.4$), the transport coefficients for the 50:50 mixture fall outside the range bounded by the two pure species, a behavior not seen at lower density ($\Phi = 0.2$). This non-linearity underscores the importance of particle correlations in dense mixtures and their impact on transport properties.

As demonstrated in~\cite{Boyle_2024}, even modest amounts of dopant or secondary species can dramatically alter electron transport in noble liquid systems, sometimes enhancing and other times suppressing mobility and diffusion compared to the pure fluids. This opens new possibilities for tuning detector media by carefully selecting admixture compositions to optimize signal response, timing, or gain.

Although coherent elastic scattering dominates at low electron energies, a complete description of transport in mixtures also requires accounting for how dopants affect polarization screening and bulk potential contributions. Extending the framework developed in Section~\ref{sub:potentials} to multi-component systems is a subject of ongoing work, and will be essential for predictive modeling of next-generation doped noble-liquid detectors.

\subsection{Electron transport across the gas-liquid interface}\label{sub:interface} 

Dual-phase TPCs take advantage of both the liquid and gas phases of noble elements, and are central to many state-of-the-art detectors~\cite{EFCA2021,Baudis2024}. In these systems, a well-defined gas-liquid interface is intrinsic to the detector geometry, typically formed by maintaining the liquid at its boiling point through controlled vapor-condensation mechanisms. 

In a typical dual-phase detector, ionization electrons generated in the liquid drift under a modest electric field until they reach the interface. At this boundary, they are extracted into the vapor phase, where higher field strengths are used to generate secondary scintillation light or avalanche multiplication via amplification grids. The efficiency and fidelity of this extraction process are crucial for detector sensitivity, timing resolution, and energy reconstruction. However, the gas-liquid interface introduces several physical complexities that are not present in homogeneous systems.

One significant complication is the likely accumulation of surface charge at the interface in both the longitudinal and transverse directions, relative to the extracting field. Such charge buildup may lead to distortions of the local electric field, inhibit electron extraction, or produce spurious signals, especially in large-scale detectors where small distortions can impact performance over long drift distances. Moreover, the microscopic structure of the interface, which is shaped by atomic-scale density gradients and thermal fluctuations, further complicates the prediction of electron behavior near this region.

Accurately modeling electron transport across the gas–liquid boundary requires more than simply adding together the transport models from each phase. While simplified approaches often treat the interface as a sharp boundary layer with idealized transfer conditions, a more complete and physically consistent treatment must account for the continuous variation in density, potential energy, and scattering environment that electrons encounter as they approach and cross the interface.

A promising step in this direction comes from recent work by ~\cite{Garland_2018}, who modeled the interface as a smooth density gradient characterized by an equilibrium profile spanning the two phases. Key to their approach was the relation of the spatial variation in $V_0$, the quasi-free electron energy, to a varying effective electric potential which through the gradient presents as a spatially inhomogeneous effective electric field. This effective field may then functionally act to inhibit electron extraction across the interface from LAr or LXe to gas, or conversely accelerate electron transport from gas into liquid. This approach is supported by molecular dynamics and MC simulations, which show that noble liquids such as LAr and LXe exhibit finite-thickness interfacial layers with well-defined thermodynamic properties. Early studies~\citep{Chapela1977,Trokhymchuk1999} employed Lennard-Jones potentials to simulate noble gas interactions and were able to reproduce key interfacial properties, including the surface tension, equilibrium vapor and liquid densities, and the interfacial width. 

These simulations provide critical inputs for transport modeling, especially for defining the spatial variation of scattering cross-sections, background potential energy, and local electron density in the interfacial region. Incorporating this information into kinetic, fluid or MC models enables a more accurate and continuous treatment of electron transport from liquid to gas. It also provides a path towards predicting extraction efficiencies, timing characteristics, and possible sources of noise originating from interfacial phenomena. As dual-phase detectors continue to scale in size and complexity, a deeper understanding of interfacial electron dynamics grounded in molecular simulation and kinetic theory will be essential for ensuring the performance and reliability of future experiments.

\FloatBarrier

\section{Summary}\label{sec:summary}

In this work, we have provided an overview of the experimental and theoretical landscape of electron transport in noble liquids, with a particular focus on applications to liquid xenon and liquid argon time projection chambers. We reviewed key swarm measurements of electron mobility and diffusion, drawing from studies spanning several decades, and highlighted the importance of these measurements for validating transport models used in modern detectors. These measurements reveal consistent trends but also emphasize the need for high-purity systems and standardized reporting of experimental conditions.

We examined the state-of-the-art in modeling low-energy electron transport in dense media, with particular emphasis on multi-term solutions to the Boltzmann equation. Central to this theoretical framework are the use of structure-modified elastic scattering cross-sections and \textit{ab initio} interaction potentials that incorporate polarization screening and bulk medium effects. Our analysis highlighted the limitations of gas-phase scaling and reinforces the need for liquid-specific scattering models that account for collective effects and self-localization of electrons. Inelastic processes, while less well characterized, are also critical, particularly exciton formation and interband transitions, which remain active areas of research due to the lack of reliable liquid-phase cross-sections.

For next-generation noble-liquid detectors, an accurate understanding of excitation processes, and consequently, scintillation and luminescence, is essential. While progress has been made, especially in modeling excimers, predicting light yields in varying field and impurity conditions remains a significant challenge. Recent proposals involving doped and mixed noble liquids open new opportunities for tuning the optical and electronic properties of the detector medium. Furthermore, the transport of electrons across the gas-liquid interface also remains an important frontier, where improved modeling can reduce systematic uncertainties and guide interface engineering in dual-phase detectors. Emerging machine learning techniques and improved datasets offer promising tools for tackling each of these complex problems.

Looking to the future, as noble-liquid detectors continue to scale to kilotonne and even multi-kilotonne active masses, the demands on predictive transport models become more acute. Precise modeling of diffusion, recombination, and ionization dynamics is essential not only for preserving signal fidelity over meter-scale drift lengths, but also for the design of optimized detector geometries, electric field configurations, and readout systems. In this context, advances in \textit{ab initio} theory, data-driven modeling, machine learning, and targeted measurements will be critical for achieving predictive, reliable descriptions of electron transport in noble liquids, capable of spanning the full range of operating conditions from sub-Td reduced fields to high-field extraction regimes.

\section*{Conflict of Interest Statement}

The authors declare that the research was conducted in the absence of any commercial or financial relationships that could be construed as a potential conflict of interest.

\section*{Author Contributions}

GB: Conceptualization, Data curation, Formal analysis, Investigation, Methodology, Project administration, Software, Validation, Visualization, Writing – original draft, Writing – review \& editing. 
NG: Conceptualization, Data curation, Formal analysis, Investigation, Methodology, Project administration, Software, Validation, Visualization, Writing – original draft, Writing – review \& editing. 
DM: Conceptualization, Data collection, Data curation, Formal analysis, Investigation, Software, Validation, Visualization, Writing – original draft, Writing – review \& editing. 
IS: Data curation, Formal analysis, Investigation, Methodology, Software, Validation, Visualization, Writing – original draft, Writing – review \& editing. 
IS: Data curation, Formal analysis, Investigation, Methodology, Software, Validation, Visualization, Writing – original draft, Writing – review \& editing. 
DB: Data curation, Formal analysis, Investigation, Methodology, Software, Validation, Visualization, Writing – review \& editing. 
SD: Data curation, Formal analysis, Funding acquisition, Investigation, Methodology, Software, Supervision, Validation, Visualization, Writing – original draft, Writing – review \& editing. 
RW: Conceptualization, Formal analysis, Funding acquisition, Investigation, Resources, Supervision, Writing – review \& editing.

\section*{Funding}
DM and RW gratefully acknowledge the financial support of the Australian  Research Council (ARC) through the Discovery Projects Scheme (DP220101480 and  DP190100696). IS, DB, and SD acknowledge the support of the Ministry of Science, Technological Development and Innovations of the Republic of Serbia and the Institute of Physics Belgrade.

\section*{Acknowledgments}
The authors wish to thank Bob McEachran for his unrivaled expertise, scattering calculations, and continued support for the development of electron-liquid transport models. The authors would also like to thank Diego Ram\'{i}rez Garc\'{i}a for valuable discussions on TPC operation and analysis. The authors acknowledge the use of ChatGPT (GPT-4-turbo, April 9th 2024 update) by OpenAI. The prompt \textit{Proofread the following scientific manuscript for spelling and grammar errors. Ensure consistency in tense, technical terminology, and punctuation. Maintain the formal academic tone, and revise the text where needed to ensure a consistent use of first-person plural narration (e.g., “we show”, “our results”). Do not rephrase unnecessarily—keep the original style and structure wherever possible} was employed for the final draft. The authors have checked the factual accuracy of the output. 

\section*{Data Availability Statement}
The datasets for the experimental measurements considered in this study can be found at \url{www.github.com/jcu-transport-physics/liquid-transport-data}. 

\bibliographystyle{Frontiers-Harvard}
\bibliography{TransportTheory,TransportData,manually_added} 
\appendix

\section{Liquid argon and xenon data}\label{appendix:data}

The following tables present details of the experimental measurements for electrons in LAr and LXe, compiled in the data repository described in Section~\ref{sec:Swarms}.

\newpage
\phantom{spacer}
\begin{sidewaystable}[p]
\centering
\begin{tabular}{@{} r r r r 
                  r r r r
                          @{}}
\toprule
$T$ & $N$ & $P$  & Ref. &
$T$ & $N$ & $P$  & Ref. \\\relax
[K] & [m$^{-3}$] & [kPa] & &
[K] & [m$^{-3}$] & [kPa] & \\
\midrule
\addlinespace
\multicolumn{4}{c}{$\mu N\left(E/N\right)$} &
\multicolumn{4}{c}{$\mu\left(E\right)$} \\

 87.0 & 2.0e28 & - & \citep{Shinsaka1988} &
 85.0 & - & 608.0 & \citep{Schnyders1966} \\

 87.0 & 2.1e28 & - & \citep{Huang1981} &
 85.0 & - & - & \citep{Gushchin1982} \\

 94.0 & 2.0e28 & - & \citep{Huang1981} &
 85.0 & - & - & \citep{Halpern1967} \\

 120.0 & 1.8e28 & - & \citep{Huang1981} &
 85.0 & - & - & \citep{Miller1968} \\
 
 125.8 & 1.7e28 & - & \citep{Eibl1990} &
 86.8 & - & - & \citep{Kalinin1996} \\

 134.4 & 1.5e28 & - & \citep{Eibl1990} &
 87.0 & - & - & \citep{Buckley1989} \\

 137.0 & 1.5e28 & - & \citep{Huang1981} &
 87.0 & - & - & \citep{Pruett1967} \\

 138.6 & 1.5e28 & - & \citep{Eibl1990} &
 87.0 & - & - & \citep{Walkowiak2000} \\

 141.7 & 1.4e28 & - & \citep{Eibl1990} &
 87.0 & - & - & \citep{Yoshino1976} \\

 144.2 & 1.3e28 & - & \citep{Eibl1990} &
 87.0 & - & - &\citep{Abratenko2021} \\

 145.0 & 1.3e28 & - & \citep{Huang1981} &
 87.7 & - & - & \citep{Abi2020} \\

 145.2 & 1.3e28 & - & \citep{Eibl1990} &
 89.8 & - & - & \citep{Walkowiak2000} \\

 146.0 & 1.3e28 & - & \citep{Lamp1994} &
 90.0 & - & - & \citep{Williams1957} \\

 146.8 & 1.2e28 & - & \citep{Eibl1990} &
 90.1 & - & 709.3 & \citep{Schnyders1966} \\

 147.0 & 1.2e28 & - & \citep{Huang1981} &
 93.9 & - & - & \citep{Walkowiak2000} \\

 147.6 & 1.2e28 & - & \citep{Eibl1990} &
 100.0 & - & - & \citep{Gushchin1982} \\

 148.0 & 1.2e28 & - & \citep{Eibl1990} &
 120.0 & - & - & \citep{Gushchin1982} \\

 149.0 & 1.1e28 & - & \citep{Huang1981} &
 130.0 & - & - & \citep{Gushchin1982} \\

 148.4 & 1.1e28 & - & \citep{Lamp1994} &
 145.0 & - & 4590.0 & \citep{Schnyders1966} \\

 148.4 & 1.2e28 & - & \citep{Eibl1990} &
 - & - & - & \citep{Malkin1951} \\

 149.2 & 1.1e28 & - & \citep{Eibl1990} &
 & & & \\

 150.0 & 1.0e28 & - & \citep{Huang1981} &
\multicolumn{4}{c}{$\mu\left(P\right)$} \\

 150.1 & 1.0e28 & - & \citep{Eibl1990} &
 100.3 & - & - & \citep{Schnyders1966} \\

 150.5 & 9.8e27 & - & \citep{Eibl1990} &
 90.1 & - & - & \citep{Schnyders1966} \\

 150.8 & 9.2e27 & - & \citep{Eibl1990} &
 105.3 & - & - & \citep{Schnyders1966} \\

 151.0 & 8.1e27 & - & \citep{Huang1981} &
 111.5 & - & - & \citep{Schnyders1966} \\

 151.0 & 8.1e27 & - & \citep{Eibl1990} &
 145.0 & - & - & \citep{Schnyders1966} \\

 151.0 & 8.6e27 & - & \citep{Eibl1990} \\
 & & & \\
 
 & & & &
 \multicolumn{4}{c}{$D_T/\mu\left(E/N\right)$} \\
 
 \multicolumn{4}{c}{$\mu\left(E/N\right)$} &
 89.0 & - & - & \citep{Li2016} \\
 
 89.0 & - & - & \citep{Li2016} &
 - & 2.1e28 & - & \citep{Shibamura1979} \\

 & & & &
 & & & \\
 
 \multicolumn{4}{c}{$D_L\left(E\right)$} &
 \multicolumn{4}{c}{$D_T\left(E/N\right)$} \\
 
 87.0 & - & - & \citep{microboone2021} &
 - & - & - & \citep{Doke1982} \\
\bottomrule
  \end{tabular}
  \caption{List of all LAr data included in the database. Missing data is indicated by a hyphen.}
\end{sidewaystable}

\newpage
\phantom{spacer}
\begin{sidewaystable}[p]
\centering
\begin{tabular}{@{} r r r r 
                  r r r r
                          @{}}
\toprule
$T$ & $N$ & $P$  & Ref. &
$T$ & $N$ & $P$  & Ref. \\\relax
[K] & [m$^{-3}$] & [kPa] & &
[K] & [m$^{-3}$] & [kPa] & \\
\midrule
\addlinespace
\multicolumn{4}{c}{$\mu\left(E\right)$} &
\multicolumn{4}{c}{$\mu N\left(E/N\right)$} \\

 162.0 & - & - & \citep{Njoya2020} &
 163.0 & 1.4e28 & - & \citep{Huang1978} \\

 163.0 & - & - & \citep{Miller1968} &
 177.0 & 1.3e28 & 200.0 & \citep{Thieme2022} \\

 165.0 & - & - & \citep{Gushchin1982} &
 184.0 & 1.3e28 & - & \citep{Baudis2018} \\

 165.0 & - & - & \citep{Yoshino1976} &
 216.0 & 1.2e28 & - & \citep{Huang1978} \\

 167.0 & - & - & \citep{Albert2017} &
 278.0 & 8.6e27 & - & \citep{Huang1978} \\

 173.3 & - & 163.7 & \citep{Baur2023} &
 288.0 & 6.8e27 & - & \citep{Huang1978} \\

 173.3 & - & 163.7 & \citep{Baur2023} &
& & & \\

 173.4 & - & 168.0 & \citep{Jorg2022} &
\multicolumn{4}{c}{$D_L\left(E\right)$} \\

 173.4 & - & 168.0 & \citep{Jorg2022} &
 162.0 & - & - & \citep{Njoya2020} \\

 177.0 & - & - & \citep{Aprile2019} &
 177.6 & - & 205.0 & \citep{Baudis2023} \\

 177.6 & - & 205.0 & \citep{Baudis2023} &
 183.2 & - & - & \citep{Hogenbirk2018} \\

 182.0 & - & - & \citep{Aprile2012} &
& & & \\

 183.2 & - & - & \citep{Hogenbirk2018} &
\multicolumn{4}{c}{$D_T\left(E\right)$} \\

 200.0 & - & - & \citep{Gushchin1982} &
 167.0 & - & - & \citep{Albert2017} \\

 230.0 & - & - & \citep{Gushchin1982} &
& & & \\

 - & - & - & \citep{Swan1962} &
\multicolumn{4}{c}{$D_T\left(E/N\right)$} \\

& & & &
 - & 1.4e28 & - & \citep{Doke1982} \\
\bottomrule
  \end{tabular}
  \caption{List of all LXe data included in the database. Missing parameters are indicated by a hyphen.}
\end{sidewaystable}

\end{document}